\documentclass[a4paper,11pt]{article}
\usepackage[utf8]{inputenc}
\pdfoutput=1 

\usepackage{jcappub} 
\usepackage{macros}

\usepackage{parskip}

\usepackage[dvipsnames]{xcolor}
\usepackage{graphicx}
\usepackage{subfig}

\usepackage[T1]{fontenc} 

\newcommand{\arh}{{a}_{\mathrm{RH}}}

\newcommand{\rh}[1]{{#1}_{\mathrm{RH}}}
\newcommand{\gs}[1]{g_{*}(#1)}
\newcommand{\gss}[1]{g_{*S}(#1)}

\newcommand{\ths}{T_{\rm hs}}
\newcommand{\beqa}{\begin{eqnarray}}
\newcommand{\eeqa}{\end{eqnarray}}
\newcommand{\lfs}{\lambda_{\rm fs}}
\newcommand{\kfs}{k_{\rm fs}}
\newcommand{\aeq}{a_{\rm eq}}
\newcommand{\rfs}{R_{\rm fs}}
\newcommand{\seq}{\sigma_{\rm eq}}
\newcommand{\fb}{f_{\rm b}}
\newcommand{\dc}{\delta_{\rm c}}
\usepackage[numbers]{natbib}
\usepackage{bm}
\let\vec\bm

\newcommand{\diff}{\ensuremath{\mathrm{d}}}
\newcommand{\e}{\mathrm{e}}
\newcommand{\I}{\mathrm{i}}
\newcommand{\p}{\prime}

\usepackage{amsmath}
\usepackage{multirow}

\title{\boldmath Simulations of Gravitational Heating Due to Early Matter Domination}
\author[a,b,c,1]{Himanish Ganjoo\note{Corresponding author.}}
\author[d]{and M. Sten Delos}

\affiliation[a]{Department of Physics, North Carolina State University, Raleigh, NC 27695, USA}
\affiliation[b]{Department of Physics and Astronomy, University of Waterloo, Waterloo, ON, N2L 3G1, Canada}
\affiliation[c]{Perimeter Institute of Theoretical Physics, 31 Caroline St. N., Waterloo, ON, N2L 2Y5, Canada}
\affiliation[d]{Max Planck Institute for Astrophysics, Karl-Schwarzschild-Str. 1, 85748 Garching, Germany}

\emailAdd{hganjoo@ncsu.edu}

\abstract{
In cosmologies with an early matter-dominated era (EMDE) prior to Big Bang nucleosynthesis, the boosted growth of small-scale matter perturbations during the EMDE leads to microhalo formation long before halos would otherwise begin to form. For a range of models, halos can even form during the EMDE itself. These halos would dissipate at the end of the EMDE, releasing their gravitationally heated dark matter and thereby imprinting
a free-streaming cut-off on the matter power spectrum.
We conduct the first cosmological $N$-body simulations of the formation and evaporation of halos during and after an EMDE.
We show that in these scenarios, the free-streaming cut-off after the EMDE can be predicted accurately from the linear matter power spectrum. Although the free streaming can erase much of the EMDE-driven boost to density perturbations, we use our findings to show that the (re-)formation of halos after the EMDE nevertheless proceeds before redshift $\sim 1000$. Early-forming microhalos are a key observational signature of an EMDE, and our prescription for the impact of gravitational heating will allow studies of the observational status and prospects of EMDE scenarios to cover a much wider range of parameters.

}

\begin{document}
\raggedbottom
\maketitle
\setlength{\abovedisplayskip}{4pt}
\setlength{\belowdisplayskip}{4pt}

\section{Introduction}

The history of the universe between a purported period of inflationary expansion \cite{infl1,infl2,infl3} and Big Bang Nucleosynthesis (BBN) is poorly understood. Standard cosmology assumes that radiation dominated the Universe between the end of inflation and the onset of the last matter-dominated epoch. However, a wide variety of models predict alternative histories of the early universe \cite{fts-rev}, including a possible early matter-dominated era (EMDE). EMDEs can arise due to oscillating string moduli fields whose energy density dominates the early universe \cite{moduli_emde} or in theories in which long-lived hidden-sector particles become the dominant component of the Universe \cite{chen_emde,zhang,hid4,hid5,codm1,codm2}. Such scenarios also include cannibal domination \cite{cannibal_big}.
Current observations only require that the heavy species that drives an EMDE decay into Standard Model (SM) radiation by a reheat temperature $\rh{T}$ larger than a few MeV \cite{2015PhRvD..92l3534D,2019JCAP...12..012H}.

While a potentially weak SM coupling or high mass could leave this early matter species inaccessible to terrestrial experiments, the EMDE can leave a detectable gravitational imprint on the late-time Universe.
Subhorizon matter perturbations grow linearly with the scale factor during an EMDE, leading to a small-scale enhancement to the matter power spectrum that can persist in the dark matter after the end of the EMDE \cite{emde2,emde1,emde3,ae15}. This enhancement results in the formation of nonlinear structures much earlier than in standard cosmological scenarios \cite{emde2,emde1,ae15}. While these small-scale dark matter microhalos do not appreciably change the large-scale structure of the universe, their dense cores reside as substructure in current-day halos and boost the dark matter annihilation rate \cite{ae15,aew16,blanco19,sten_gr}. In addition, these structures can be detected via their effects on pulsar timing \cite{Ramani:2020hdo,pta1,pta_sten} or through their microlensing of stars \cite{caustic1,caustic2,caustic3,obs_blinov}.

If the EMDE is sufficiently long and the dark matter sufficiently cold, bound microhalos can even form and accrete dark matter during the EMDE itself. Such a scenario has been previously considered in the context of hidden-sector dark matter models \cite{blanco19} but is also a likely outcome for more minimal dark matter models (such as the supersymmetric models discussed by Ref.~\cite{aew16}; see Ref.~\cite{sten_gr}).
As the dominant particles decay, these early-forming structures evaporate, and the dark matter particles residing in them shoot out at boosted speeds in random directions.
The free streaming that results from this gravitational heating potentially erases some of the boosted small-scale power, thereby suppressing subsequent microhalo formation \cite{blanco19,barenboim_rem}. In order to predict the observational consequences of these scenarios, it is necessary to understand the nature of this suppression.

In this paper, we carry out the first cosmological $N$-body simulations of the formation of structure during an EMDE and its subsequent evaporation as the EMDE ends.
This process was previously studied by Refs.~\cite{blanco19,barenboim_rem} using analytic approximations.
We find that the impact of gravitational heating can be accurately modeled as a free-streaming cut-off to the matter power spectrum, as suggested by Ref.~\cite{blanco19}.
Moreover, even though gravitational heating is a nonlinear process, the cut-off scale is related in a simple way to the matter power spectrum calculated in linear theory, which has been studied in detail for EMDEs (e.g., Refs.~\cite{emde2,hg23}). Consequently, the matter power spectrum long after reheating can be predicted straightforwardly from such analytic results.
As a demonstration, we show that the suppression of structure due to gravitational heating is sufficiently weak that dark matter microhalos are expected to begin forming either before or not long after matter-radiation equality, even in scenarios with an arbitrarily large degree of structure formation during the EMDE.

This paper is organized as follows. In Section \ref{background}, we review the evolution of the homogeneous background and density perturbations during an EMDE. Section \ref{heating} explains the formation of nonlinear structures during an EMDE and the idea of gravitational heating due to their evaporation. In Section \ref{simulations}, we describe our suite of $N$-body simulations and show qualitatively how halos form and subsequently dissipate. Section \ref{cut-off} analyzes the free-streaming cut-off to the power spectrum that arises in the simulations after reheating and develops a connection between the scale of this cut-off and the linear power spectrum during the EMDE. In Section~\ref{latestructure}, we use this connection to explore how gravitational heating impacts the (re-)formation of structure long after reheating, particularly focusing on how the time of microhalo formation is affected. Finally, Section~\ref{summary} summarizes our results and looks at future prospects for extending and applying them. This paper uses natural units throughout, in which $c = \hbar = 1$.

\section{Background and Perturbation Evolution}
\label{background}

\subsection{The Homogeneous Background}

Although we expect the results of this work to be applicable to EMDEs in general, for concreteness we adopt the class of hidden-sector models considered in Ref.~\cite{hg23} (hereafter G23). These models consist of a universe with three fluids: the dark matter ($X$), the $Y$ particles that drive the EMDE, and the SM radiation (denoted by the subscript $R$). The $X$ and $Y$ particles reside in a hidden sector kinetically decoupled from the SM, with temperature $\ths$, while the temperature of the SM is $T$. The $X$ particles are nonrelativistic with their energy density evolving with scale factor as $\rho_X \propto a^{-3}$. The $Y$ particles are initially relativistic but become nonrelativistic as $\ths$ decreases. The nonrelativistic $Y$ particles come to dominate the energy density of the Universe and cause an epoch of early matter domination, but they decay into SM radiation with a rate $\Gamma$. Although we do not appeal to any dark matter-SM interaction in this work, we note that in the models, the $Y$ particles are the lightest in the hidden sector and act as mediators between the heavier $X$ and the Standard Model; dark matter annihilates via $XX \rightarrow YY$.

The equations for the homogeneous energy densities of the three components are: \begin{subequations} \label{eqback}
\begin{align}
    \dot{\rho}_Y + 3H(1 + w_Y)\rho_Y &= - \Gamma mn_Y; \label{ry} \\ 
    \dot{\rho}_R + 4H\rho_R &=  \Gamma mn_Y; \label{rr} \\ 
    \dot{\rho}_X + 3H\rho_X &= 0, \label{rx}
    \end{align}
\end{subequations} where overdots denote $d/dt$ and $H \equiv \dot{a}/a$. In Eq.~(\ref{ry}), $n_Y$ is the number density of $Y$ particles while $w_Y$ is their time-varying equation of state parameter, the ratio of their pressure to their density. As $w_Y$ falls from 1/3 to 0, the $Y$ particles transition from being relativistic to nonrelativistic. The terms on the RHS of Eqs.~(\ref{ry}) and (\ref{rr}) depend on $mn_Y$ instead of $\rho_Y$ because the longer lab-frame lifetimes of faster particles compensate for the higher energies released by their decays \cite{cannibal_big}. For details of the methods used to numerically solve for the evolution of the hidden sector temperature and the density and pressure of the $Y$ particles, we refer the reader to G23.

\begin{figure}[h!]
\centering
\includegraphics[scale=1.2]{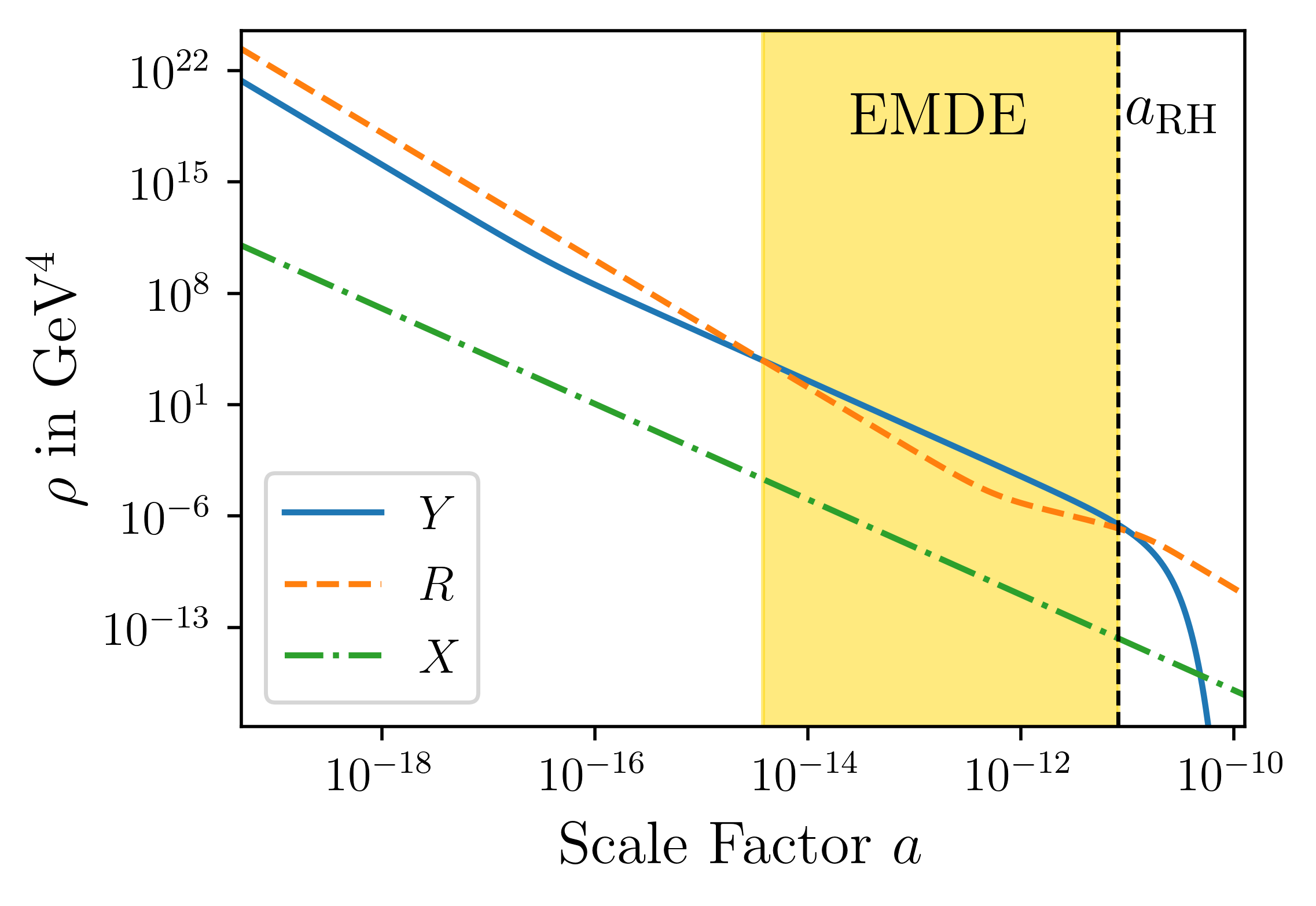}
\caption{The background evolution of the energy densities of the $Y$ particles, SM radiation ($R$) and dark matter ($X$) as a function of scale factor. The yellow shaded region shows the EMDE. The particular scenario exhibited here is one in which the SM radiation initially dominates ($\eta>1$); scenarios are also possible in which relativistic $Y$ particles initially dominate ($\eta<1$).}
\label{fig:back}
\end{figure}

Figure \ref{fig:back} shows the time evolution of the background densities of the three fluids, obtained by solving Eqs.~(\ref{eqback}). Equation~(\ref{rr}) shows that the SM radiation density drops as $\rho_R \propto a^{-4}$ when when $\Gamma mn_Y \ll H \rho_R$. When $\Gamma mn_Y$ becomes comparable to $H \rho_R$, $\rho_R \propto a^{-3/2}$ due to the entropy injection from the decay of the $Y$ particles into the visible sector. After the $Y$ particles decay away, $\rho_R \propto a^{-4}$ again. 

At the start of our calculations, $\rho_Y \propto a^{-4}$ because the $Y$ particles are relativistic. As $\ths$ drops below the $Y$ particle mass ($m$), the $Y$ particles become nonrelativistic. 
This transition can be modelled as a broken power law with a sharp pivot point at the scale factor $a_p = b a_i T_{\rm hs,i} / m$, where $b=2.70$ for bosonic $Y$ particles and $3.15$ for fermionic $Y$ particles \cite{hg23},
and the subscript $i$ indicates any time when the $Y$ particles are relativistic. When the $Y$ particles become nonrelativistic, $\rho_Y \propto a^{-3}$ until $\Gamma/H \sim 1$. When $\rho_Y$ exceeds $\rho_R$, the universe becomes matter-dominated. This phase is shown by the yellow shaded region in Figure \ref{fig:back}. When $\Gamma / H$ becomes comparable to unity, the decay of the $Y$ particles begins to significantly affect their abundance, and their comoving number density falls rapidly. Radiation domination starts shortly thereafter. This process is called \textit{reheating}, and, although it does not happen instantaneously, it is conventional to define the reheat temperature via the relation \begin{equation}\label{gamma}
    \Gamma \equiv \sqrt{\frac{8 \pi G}{3} \frac{\pi^2}{30} \gs{\rh{T}} \rh{T}^4} \,,
\end{equation} which equates $\Gamma$ to what the Hubble rate would be in a purely radiation-dominated universe at temperature $\rh{T}$. Another quantity that is useful to define is the scale factor $\arh$ of reheating, given by \begin{equation}\label{arha0}
    \frac{\rh{a}}{a_0} = \frac{1}{1.02} \left[ \frac{\gss{T_0}}{\gss{0.204 \rh{T}}} \right]^{\frac{1}{3}} \left[ \frac{T_0}{\rh{T}} \right] ,
\end{equation} where $a_0$ and $T_0$ are respectively the scale factor and the radiation temperature today \cite{hg23}. Note that both $\rh{T}$ and $\rh{a}$ are defined quantities and $T(\rh{a}) \neq \rh{T}$.

The behavior of the homogeneous background universe is controlled by three parameters. The ratio $\eta$ of the SM and $Y$ energy densities when the $Y$ particles are relativistic controls which component dominates the universe before the EMDE. The $Y$ particle mass $m$ decides when the $Y$ particles become nonrelativistic. Together, $\eta$ and $m$ determine when $\rho_Y$ exceeds $\rho_R$. For a fixed $\eta$, a higher value of $m$ means the $Y$ particles become nonrelativistic earlier and the EMDE begins earlier. For the same value of $m$, a higher value of $\eta$ means the EMDE begins later. In addition to these, $\rh{T}$ determines when the EMDE ends by setting the decay rate of the $Y$ particles {\textemdash} a smaller value of $\rh{T}$ for the same $\eta$ and $m$ means that reheating happens later and the EMDE is longer. 

The case shown in Fig. \ref{fig:back} has $\eta=100$, which implies that SM radiation dominates the energy budget of the universe before the EMDE. There can be cases with $\eta<1$, in which $\rho_Y > \rho_R$ initially and the EMDE begins when the $Y$ particles become nonrelativistic. 

\subsection{Perturbation Evolution and Power Spectra}

Subhorizon matter perturbations evolve linearly with the scale factor in a matter-dominated era, as opposed to the logarithmic evolution seen during radiation domination \cite{ae15}. As a result, in cosmologies with an EMDE, small-scale modes which enter the horizon during or before the EMDE are enhanced in amplitude compared to cases without an EMDE. This effect corresponds to an enhancement of the matter power spectrum on small scales. Additionally, the relativistic pressure of the $Y$ particles acts against gravity and suppresses the growth of $Y$ density perturbations $\delta_Y$. Modes which enter the horizon when the $Y$ particles have significant pressure experience inhibited growth until the $Y$ particle becomes nonrelativistic and the pressure subsides. This effect leads to a small-scale cut-off in the power spectrum of $\delta_Y$. 

During the EMDE, the $Y$ particles are the dominant component in the universe and they cluster to form gravitational wells. The $X$ particles fall into these gravitational centers. As a result, the $X$ density perturbations $\delta_X$ track $\delta_Y$ and the dark matter ($X$) power spectrum inherits the shape and amplitude of the $\delta_Y$ power spectrum, including the enhancement due to the EMDE and the small-scale cut-off resulting from the relativistic pressure of the $Y$ particles. 

\begin{figure}[h!]
    \centering
    \includegraphics[width=0.96\textwidth]{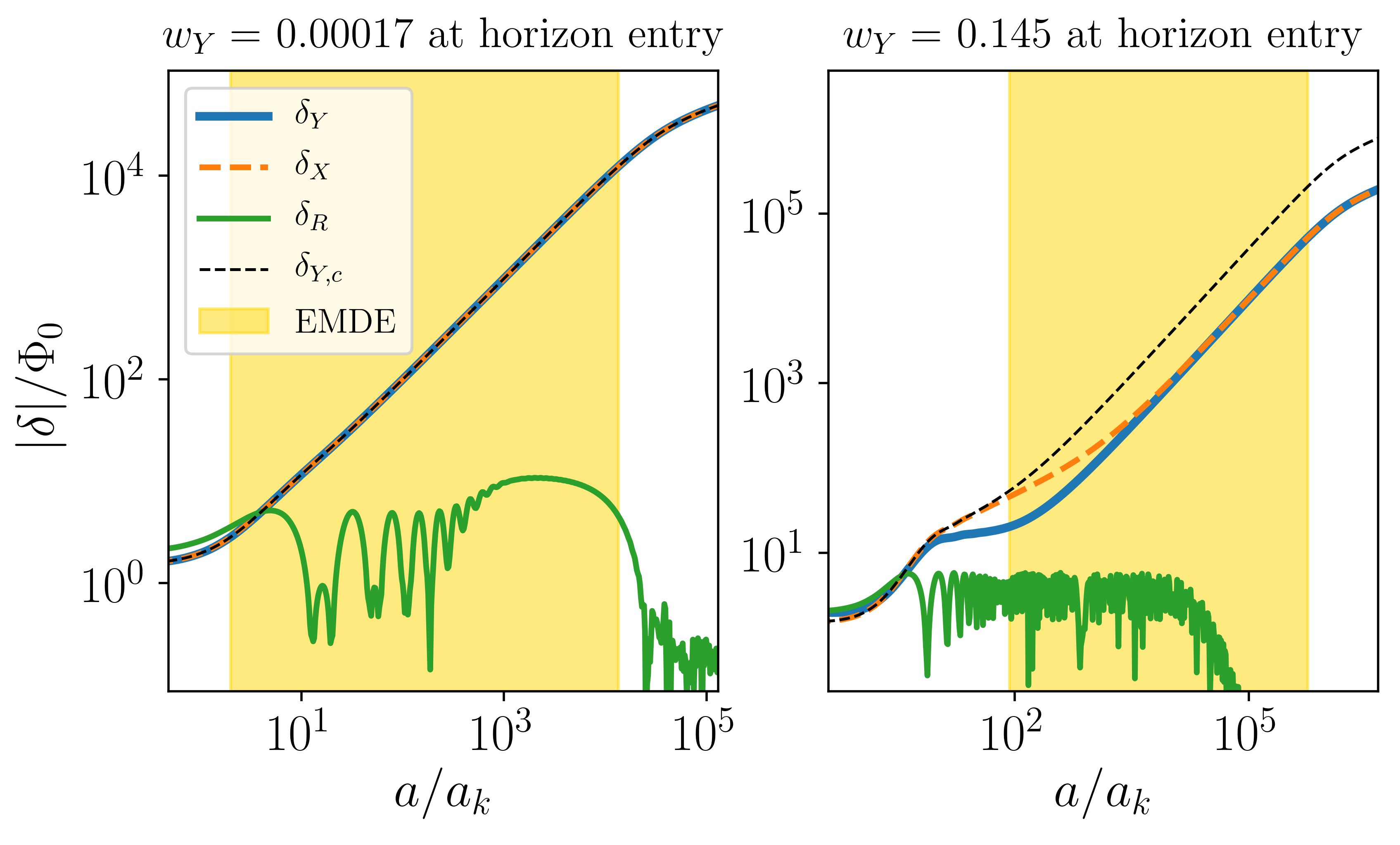} 
    \caption{The evolution of perturbations for two modes, normalized to the primordial potential and plotted as a function of $a/a_k$, where $a_k$ is the scale factor of horizon entry for the mode. The mode in the left panel enters the horizon after the $Y$ particles have become pressureless ($T_{\rm hs}/m =0.00018$ at horizon entry, corresponding to equation of state parameter $w_Y=0.00017$); there is no suppression of $\delta_Y$ for this mode. The mode in the right panel enters the horizon when the $Y$ particles have significant pressure ($T_{\rm hs} /m = 0.21$ at horizon entry, corresponding to $w_Y=0.145$) because of which $\delta_Y$ is suppressed compared to $\delta_{Y,c}$, the $Y$ density perturbation if the $Y$ particles are nonrelativistic. Dark matter perturbations $\delta_X$ eventually track $\delta_Y$ in both cases.} \label{fig:modes}   
\end{figure}

Figure \ref{fig:modes} (taken from G23) illustrates the time evolution of two perturbation modes in Newtonian gauge, showing $\delta_X$ and $\delta_Y$ scaled by $\Phi_0$, the primordial metric perturbation in a radiation-dominated universe. It also shows the evolution of $\delta_{Y,c} / \Phi_0$, the $Y$ density perturbation if the $Y$ particles were pressureless.  The left panel shows the evolution of the density perturbations for a mode which enters the horizon when the $Y$ particles are nearly pressureless, as evidenced by the equation of state $w_Y$ at horizon entry being 0.00017. In this case, $\delta_X$, $\delta_Y$ and $\delta_{Y,c}$ evolve like a cold dark matter perturbation. The right panel shows the evolution of the same three for a mode which enters the horizon when $w_Y = 0.18$ and the $Y$ particles have significant relativistic pressure. Here, the growth of $\delta_Y$ is initially suppressed compared to $\delta_{Y,c}$, before the pressure subsides and $\delta_Y$ starts growing linearly with scale factor during the EMDE. In addition, $\delta_X$ starts tracking $\delta_Y$ after some time has elapsed in the EMDE, on account of falling into the gravitational wells created by the $Y$ particles. 

\begin{figure}[h!]
\centering
\includegraphics[scale=1.2]{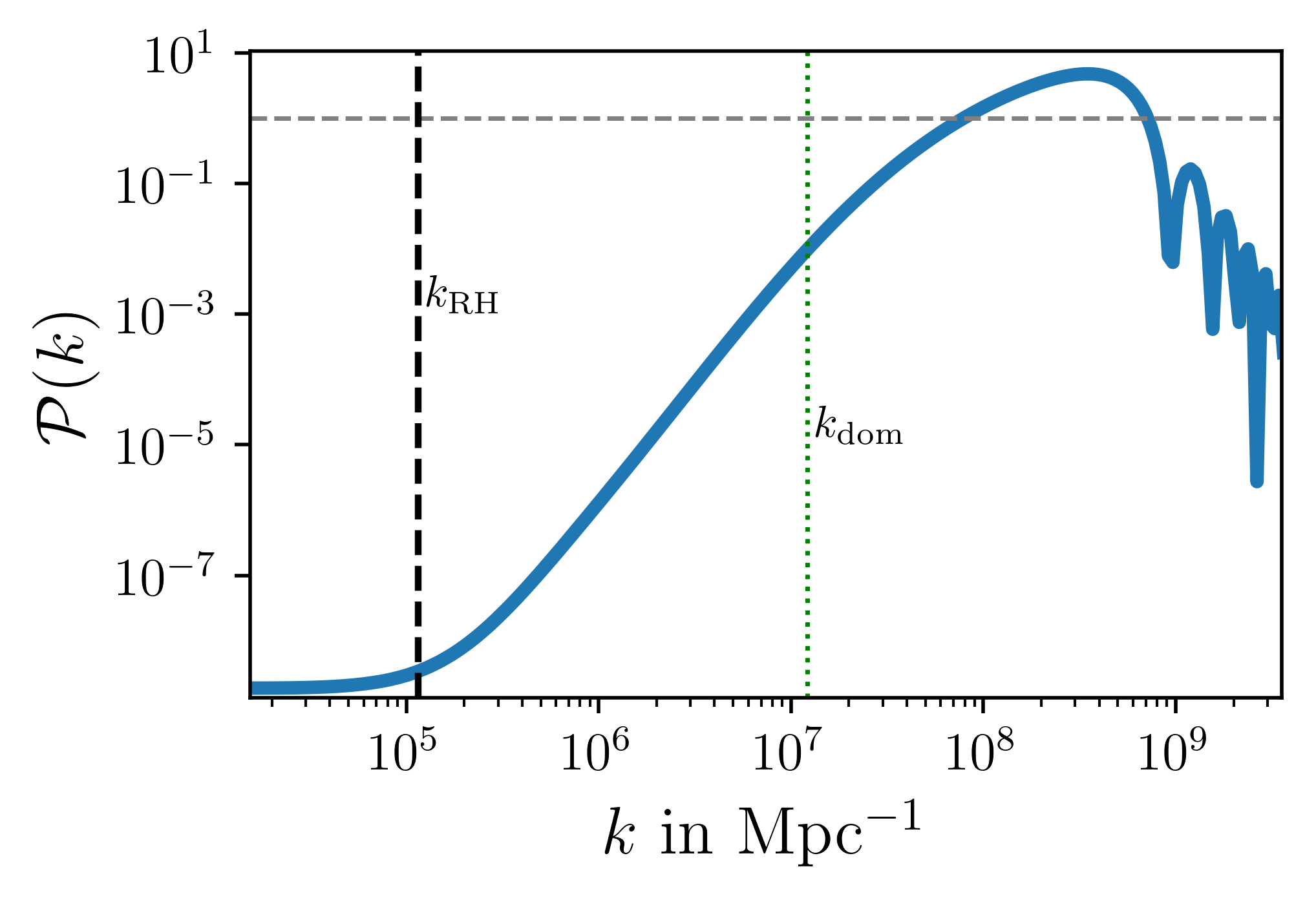}
\caption{The dimensionless power spectrum $\mathcal{P}(k)\equiv[k^3/(2\pi^2)]P(k)$ of dark matter density perturbations, evaluated in linear theory at $\rh{a} = 1.64 \times 10^{-11}$ for a cosmology with $\rh{T} = 10$ MeV, $\eta = 100$ (implying that SM radiation initially dominates), and a bosonic $Y$ particle of mass $m = 1$ TeV that dominates the energy density of the universe during the EMDE. The vertical dashed line shows $\rh{k}$, roughly the wavenumber corresponding to the horizon at the end of the EMDE. The dotted line marks the wavenumber that enters the horizon at the onset of the EMDE.}
\label{fig:ps}
\end{figure}

The exact nature of the effect of the $Y$ particle pressure on the matter power spectrum was worked out in G23 using a custom Boltzmann solver that incorporated the effects of the varying equation of state and sound speed of the $Y$ particles into the perturbation equations for the three fluid densities and velocities. Figure \ref{fig:ps} shows one example solution, showing the dimensionless $X$ power spectrum $\mathcal{P}(k)\equiv[k^3/(2\pi^2)]P(k)$ evaluated in linear theory: \begin{equation}\label{power}
    \mathcal{P}(k) \equiv A_s \left(\frac{k}{k_0} \right)^{n_s-1} \left( \frac{4}{9} \right) \left(\frac{\delta_X(k)}{\Phi_0} \right)^2,
\end{equation} where $A_s = 2.2 \times 10^{-9}$, $k_0 = 0.05$ Mpc$^{-1}$, and $n_s = 0.965$ are taken from the Planck 2015 results \cite{planck15}.
The first two factors in Eq.~(\ref{power}) are the power spectrum of primordial curvature perturbations $\zeta$, and the remaining factors translate this into the matter power spectrum (since the primordial gravitational potential $\Phi_0 = -2\zeta / 3$).
The power spectrum is shown at $\rh{a} = 1.64 \times 10^{-11}$ for a case with a bosonic $Y$ particle of $m = 1$ TeV, $\eta = 100$ (so that SM radiation initially dominates), and $\rh{T} = 10$ MeV. The vertical dashed line shows $\rh{k} \equiv \rh{a} \Gamma$. For modes with $k \gtrsim \rh{k}$, the power spectrum is enhanced due to the EMDE. Modes that enter the horizon during the EMDE ($\rh{k} < k < k_{\rm dom}$) grow linearly with $a$ from horizon entry until $\rh{a}$. For these modes, $\delta_X(k,\rh{a}) \propto (k/\rh{k})^2$ and $\mathcal{P}(k) \propto (k/\rh{k})^4$. Modes with $k > k_{\rm dom}$ enter the horizon in the radiation-dominated era before the EMDE and experience logarithmic growth with $a$ before linear growth during the EMDE. If the $Y$ particles were pressureless, this growth would imply that $\mathcal{P}(k) \propto [\ln (k/k_{\rm dom})]^2$ for $k > k_{\rm dom}$. This transition to logarithmic shape at $k_{\rm dom}$ is apparent in Fig. \ref{fig:ps}. In contrast, for a case with initial $Y$ domination ($\eta<1$), $\mathcal{P}(k) \propto (k/\rh{k})^4$ for all $k > \rh{k}$ down to the cut-off scale arising due to the $Y$ particle pressure.

Due to the pressure of the $Y$ particles, the growth of modes with $k \gtrsim 10^8$~Mpc$^{-1}$ in Fig.~\ref{fig:ps} is suppressed. This suppression leads to a characteristic peak in the power spectrum. For scales smaller than the peak scale, the amplitude of $\mathcal{P}(k)$ displays an oscillating pattern with a decaying envelope. This decaying amplitude is because smaller scales enter the horizon earlier and experience a longer period of suppressed growth. 

\section{Structure Formation During The EMDE and Gravitational Heating}
\label{heating}

In the scenario shown in Fig.~\ref{fig:ps}, the linearly extrapolated matter power spectrum significantly exceeds 1 (dashed horizontal line) at reheating. Consequently, nonlinear structures would have already formed during the EMDE.
According to Press-Schechter theory \cite{press-schechter}, the fraction $f_\mathrm{bound}$ of matter in halos is
\begin{equation}\label{fbound}
    f_\mathrm{bound}(a) = 2\int_{\dc/\sigma}^\infty \diff\nu \frac{\e^{-\nu^2/2}}{\sqrt{2\pi}}
    =\mathrm{erfc}\left(\frac{\dc}{\sqrt{2}\sigma(a)}\right)
\end{equation}
at the scale factor $a$, where $\sigma(a) = [\int \mathcal{P}(k,a)\diff \ln k]^{1/2}$ is the rms of the unfiltered linear density contrast field, $\dc\simeq 1.686$ is the linear threshold for spherical collapse, and $\mathrm{erfc}(x)\equiv 1-\mathrm{erf}(x)$ is the complementary error function.\footnote{In Press-Schechter theory, particles lying at a point in the initial conditions become part of a halo at some later time if the linear density contrast about that point exceeds the threshold, $\dc$, when smoothed on some scale. In Eq.~(\ref{fbound}), $f_\mathrm{bound}/2$ is the fraction of points at which the unfiltered density exceeds the collapse threshold. The factor of 2 accounts for the points for which the density only exceeds the threshold when smoothed on a nonzero scale \cite{1991ApJ...379..440B}.}
To quantify the dominance of nonlinear structure formation during the EMDE, we consider $f_\mathrm{bound}(\rh{a})$, the fraction of $Y$ particles bound in halos by the time of reheating. For a case with $\{m,\eta,\rh{T}\} = \{2 \text{ TeV},10, {20} \text{ MeV} \}$, the bound fraction is 0.805. That is, more than 80\% of the $Y$ particles are in bound structures at reheating in this case. Since the $X$ particles fall into the gravitational wells created by the $Y$ particles, we can assume a similar fraction of $X$ particles are in bound halos at this time.

Bound microhalos that form before reheating are mostly composed of the $Y$ particles, since $\rho_Y \gg \rho_X$. Around the time of reheating, as the decay of the $Y$ particles suppresses their abundance, the gravitational support of these structures vanishes. The $X$ particles that have virialized in a microhalo cannot adjust their orbits adiabatically to this evaporation of the halo's mass, so they leave the halo at boosted speeds in random directions. This gravitational heating leads to free streaming of dark matter after reheating, which suppresses density perturbations on comoving length scales smaller than the free-streaming length $\lfs$. Reference \cite{blanco19} estimated that structures formed after about $10^{-6} \rh{a}$ will dissipate at reheating, releasing their dark matter particles to stream freely. 

If the population of dark matter particles has some (peculiar) velocity dispersion $\sigma_v$, the free-streaming length is given by \begin{equation}
    \lfs (t) \sim \int^{t}_{\rh{t}} \frac{\sigma_v(t^\prime)}{a} \diff t^\prime.
\end{equation} Assuming that nonlinear growth stops at $\rh{a}$ and the velocity dispersion redshifts after that as $\sigma_v(a) = \rh{\sigma_v} \rh{a} / a$, the above integral can be recast as \begin{equation}
    \lfs (a) \sim  \rh{\sigma_v} \rh{a}\int^{a}_{\rh{a}} \frac{\diff a^\prime}{a^{\prime 3} H(a^\prime)}. 
\end{equation} During the period of radiation domination that follows the EMDE, we have $H(a) \propto a^{-2}$, where we neglect the minor influence of changes in the relativistic content. This implies $a^2 H(a) = \rh{a}^2 H(\rh{a})$, using which the integral is \begin{equation}\label{lfs}
    \lfs (a) \sim  \frac{\rh{\sigma_v}} {\rh{a} H(\rh{a})}\int^{a}_{\rh{a}} \frac{\diff a^\prime}{a^\prime} = \frac{\rh{\sigma_v}} {\rh{a} H(\rh{a})} \ln \left[\frac{a}{\rh{a}}\right].
\end{equation} Finally, defining $\kfs = \lfs^{-1}$ and using $\rh{k} = \rh{a} \Gamma = 0.96\rh{a} H(\rh{a})$,\footnote{\label{foot:H_RH}The numerical factor here corresponds to $\Gamma=H(\rh{T})=0.96 H(\rh{a})$, where $H$ is the radiation-dominated extrapolation of the Hubble rate, and follows from Eq.~(\ref{arha0}).}
where $H(\rh{a})$ is the Hubble rate in a radiation-dominated universe, we can write \begin{equation}\label{kfs}
    \frac{\kfs}{\rh{k}} \sim \frac{1}{\rh{\sigma_v}} \left(\ln \left[\frac{a}{\arh} \right] \right)^{-1}.
\end{equation}
Note that this discussion is conceptual, as $\rh{\sigma_v}$ is set by nonlinear gravitational evolution and cannot be evaluated directly (see Sec.~\ref{cut-off} for a numerically precise analysis).

This free streaming will result in a suppression of power on scales with $k \gtrsim \kfs$, leading to a fall-off in the matter power spectrum after the EMDE ends. As Eq.~(\ref{kfs}) shows, the cut-off scale grows larger ($\kfs$ decreases) with time, as the dark matter particles cover increasing distances. This growth is logarithmic during the radiation epoch but comes to a halt after matter-radiation equality, since during matter domination, the integral $\int da / [a^3 H(a)]$ converges in the infinite future.

As the free-streaming scale grows larger, the enhancement to the power spectrum resulting from the EMDE is erased up to larger scales. Yet as long as $\kfs / \rh{k} > 1$ around the time of matter-radiation equality, some of the enhancement to the power spectrum due to the EMDE remains even after the impact of gravitational heating, and microhalos can start forming much earlier than they do in standard $\Lambda$CDM scenarios. Since the density of a halo forming at $a_f$ scales as $a_f^{-3}$, these early-forming structures will have cores much denser than halos which form in standard cosmologies,
greatly boosting their annihilation signals while also increasing their susceptibility to detection by gravitational means. Accurate modeling of the free-streaming scale and the manner in which it suppresses the matter power spectrum is therefore important for developing observational constraints on these EMDE scenarios.

Reference~\cite{blanco19} previously estimated the cut-off to the power spectrum imposed by gravitational heating by considering the virial velocities of the halos that form during the EMDE. We improve upon this treatment by using cosmological $N$-body simulations, which we describe next.

\section{Simulations}
\label{simulations}

To analyze the formation of nonlinear structures during the EMDE and the impact of their subsequent evaporation on the matter power spectrum, we used a modified version of the $N$-body simulation code \textsc{GADGET-2} \cite{gadget2}. The code was changed to include SM radiation as in Refs.~\cite{sten_ucmh,sten_ucmh_prof} by adding homogeneous density terms to the expression for the Hubble rate in the time integrals that are used to calculate the drifts and kicks in the simulations. The decaying $Y$ particles are modeled in the same way at the homogeneous level, and their inhomogeneity is accounted for by scaling the efficiency of gravitational kicks proportionally to the time-varying quantity $(\bar{\rho}_Y + \bar{\rho}_X)/\bar{\rho}_X$;\footnote{More specifically, we scale the time integrand for gravitational kicks. This approach avoids introducing artifacts associated with the finite time-step resolution.} this is conceptually equivalent to scaling the simulation particle mass $M$ as
\begin{equation}
    M = \left( \frac{\bar{\rho}_Y + \bar{\rho}_X }{\bar{\rho}_X} \right) M_0,
\end{equation}
where $M_0$ is the simulation particle mass at late times. This scheme was developed for and tested in Ref.~\cite{blanco19}, and similar methods have been employed in simulations of decaying dark matter \cite{suto87,enqvist15,dakin19}. Since the $X$ and $Y$ particles are coupled gravitationally during the EMDE, we represent both species together in each simulation particle.

We used our modified code to run a suite of nine cosmological simulations. The initial conditions in each case were set by sampling a random initial density field from the power spectrum of $Y$ particles obtained using the Boltzmann solver described in G23. Simulation particles were initially on a uniform grid, and their positions and velocities were perturbed using the Zel'dovich approximation.

We began all of our simulations during the EMDE, at $a = 0.02 \rh{a}$, with the exception of two cases (YD1 and YD2), which began at the earlier time $a=0.01 \rh{a}$ and $0.005\rh{a}$, respectively. The starting times were chosen to ensure that the matter power spectrum is initially in the linear regime, as much as possible. Periodic boundary conditions were assumed and the comoving box sizes were set to half the size of the comoving horizon at the starting time. The RMS density fluctation at the scale of the mean interparticle distance, $\sigma=[\int\mathcal{P}(k,a_{\rm start})\diff\log k]^{1/2}$ in each box was less than 0.15 at the starting time, with the exception of YD2, for which it was 0.35. The comoving softening length for the simulation particles was set to 0.03 times the initial grid spacing.

\begin{table}[]
\begin{tabular}{|c|c|c|c|c|c|c|}
\hline
\textbf{Case} & \textbf{$m$ (TeV)} & \textbf{$\eta$} & \textbf{$\rh{T}$ (GeV)} & \textbf{$\rh{a}$} & \textbf{$\rh{k}$ (1/Mpc)} & Bound \% at $\rh{a}$ \\ \hline \hline
RD1           & 2                  & 70             & 0.02                    & $8.21\times 10^{-12}$          & $2.36 \times 10^5$        & 59.66               \\ 
RD2           & 2                  & 10             & 0.02                    & $8.21\times 10^{-12}$         & $2.36 \times 10^5$        & 80.54               \\ 
RD3           & 2                  & 10             & 0.05                    & $3.29\times 10^{-12}$          & $6.68 \times 10^5$        & 36.02               \\ 
YD1           & 2                  & 0.01           & 0.02                    & $8.21\times 10^{-12}$        & $2.36 \times 10^5$        & 86.72               \\ 
YD2           & 2                  & 0.01           & 0.005                   & $3.31\times 10^{-11}$          & $5.77 \times 10^4$        & 97.23               \\ 
YD3           & 2                  & 0.01           & 0.069                   & $2.37\times 10^{-12}$          & $9.51 \times 10^5$        & 32.84               \\ 
YD4           & 20                 & 0.01           & 0.5                     & $2.85\times 10^{-13}$          & $1.26 \times 10^7$        & 27.46               \\ 
RD4           & 2                  & 400            & 0.02                    & $8.21\times 10^{-12}$         & $2.36 \times 10^5$        & 14.61               \\ 
RD5           & 2                  & 100            & 0.05                    & $3.29\times 10^{-12}$     & $6.68 \times 10^5$        & 1.81                \\ \hline
\end{tabular}
\caption{List of simulations with the model parameters $m$, $\eta$ and $\rh{T}$, the values of $\rh{a}$ and $\rh{k}$, and the percentage of matter bound in halos at reheating, evaluated using Eq.~(\ref{fbound}).}
\label{tab:sim}
\end{table}

Table \ref{tab:sim} lists the hidden-sector model parameters $\{m,\eta,\rh{T}\}$ and the details of the various simulation runs. The three parameters were varied to cover cases with different amounts of predicted bound structure at $\rh{a}$ to check the impact of differing levels of structure formation on the post-reheating cut-off that arises due to gravitational heating. Our simulations go from a case with 1\% bound structure at reheating to a case with nearly all of the matter bound in halos at reheating.  The range of models is also associated with a variety of different shapes of the linear matter power spectrum during the EMDE.

\begin{figure}
\begin{tabular}{cc}
  \includegraphics[width=0.475\textwidth]{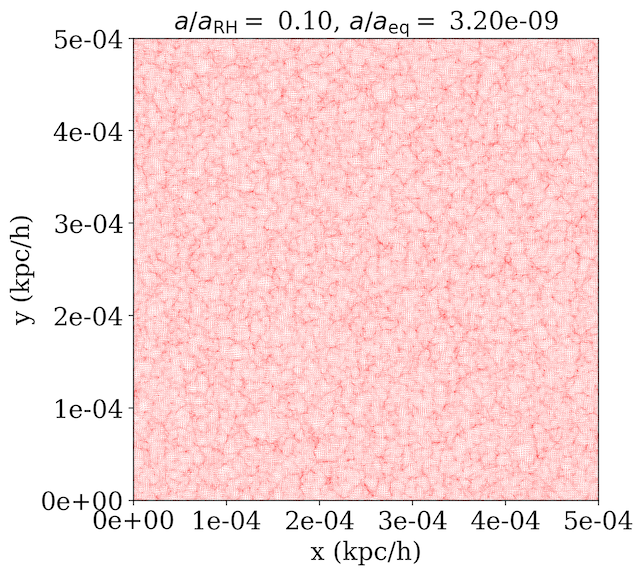} &   \includegraphics[width=0.475\textwidth]{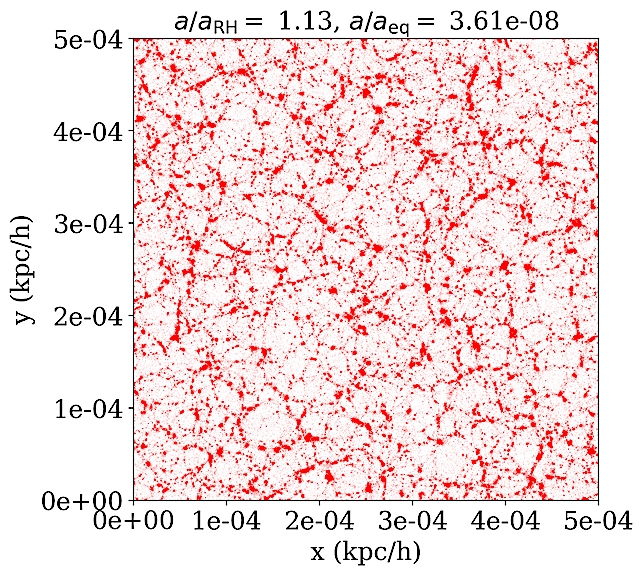} \\
 \includegraphics[width=0.475\textwidth]{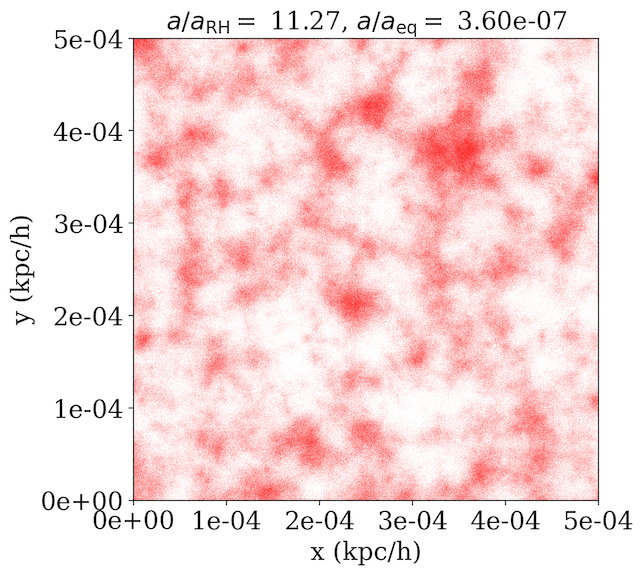} &   \includegraphics[width=0.475\textwidth]{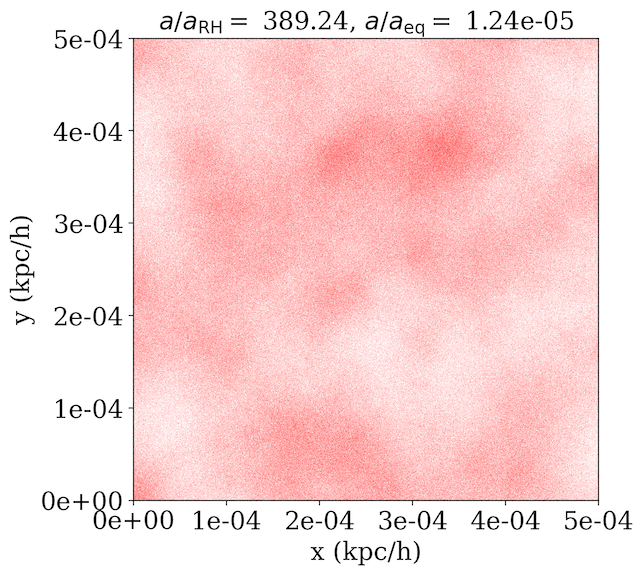} \\
\end{tabular}
\caption{Density plots of slices of thickness $4.5 \times 10^{-5}$ kpc/$h$ from the RD2 simulation box, for which $\rh{a} = 8.21 \times 10^{-12}$. The box size is $9.01 \times 10^{-4}$ kpc/$h$; the x-y plane is shown from 0 to $5 \times 10^{-4}$ kpc/$h$ . The four slices show the formation and erasure of structure as the box evolves towards and past reheating.}
\label{fig:slices}
\end{figure}

Figure \ref{fig:slices} shows slices of the RD2 simulation box at various times. The slices have a thickness of 0.05 times the box size and extend across 0.45 times the box size in the other directions; the particles are plotted as dots. The top left panel shows the box at $0.10\rh{a}$, when the perturbations in the density field are small. The top right panel shows the box just after reheating, at $1.13\rh{a}$, displaying the formation of a web of structure with clumps and filaments. The bottom left panel displays the box at $11.27 \rh{a}$. By this time, the structure formed at reheating is already being erased, with the clearly defined web visible at $1.13\rh{a}$ blurred and smoothed out. By $389.24 \rh{a}$ (bottom right panel), the structure formed during the EMDE has been almost completely wiped out. The progression of the box shown via these four snapshots illustrates the formation of structures during the EMDE and their subsequent dissipation after reheating. 

\begin{figure}[h!]
\centering
\includegraphics[scale=0.9]{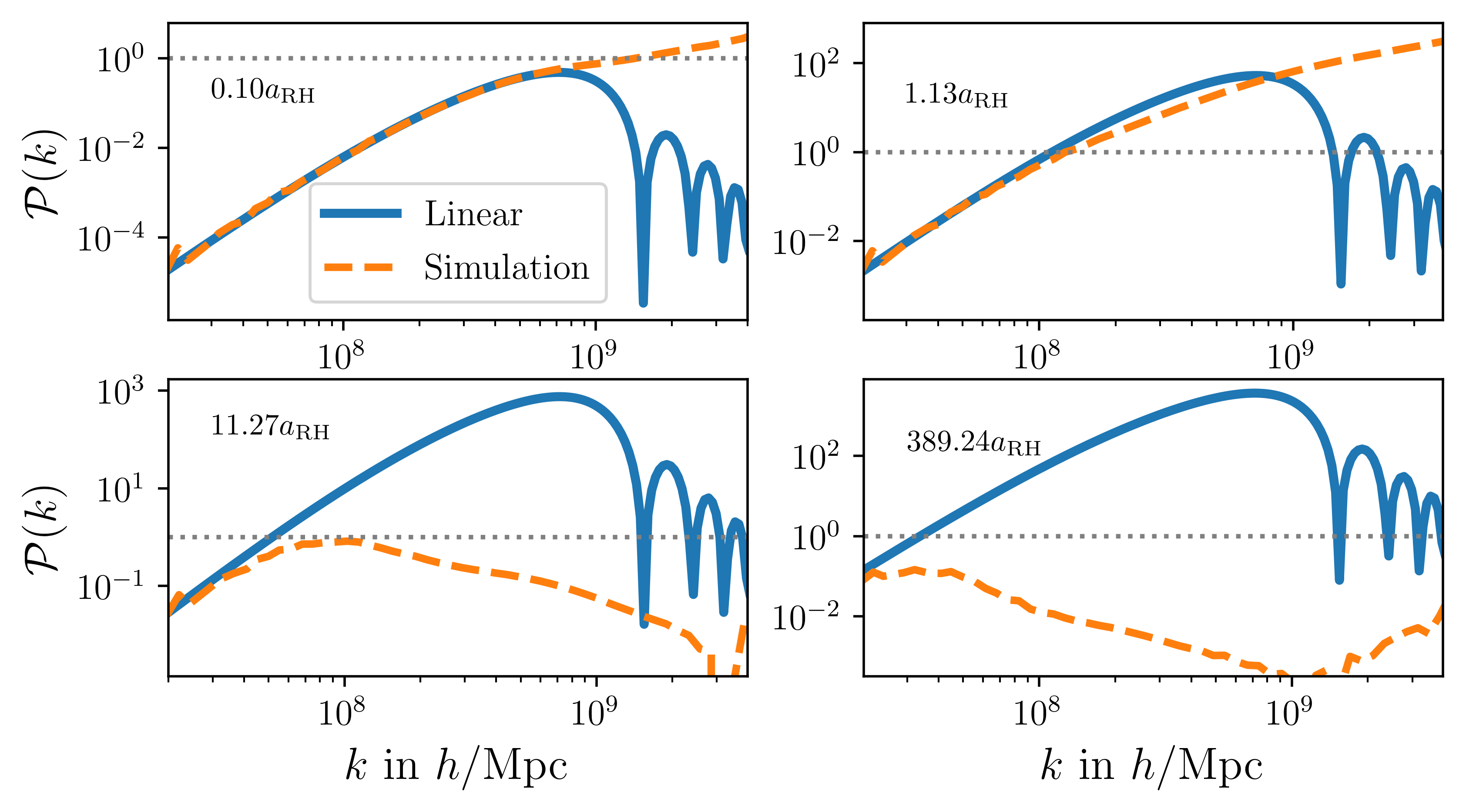}
\caption{The dimensionless power spectrum $\mathcal{P}(k)$ taken from the RD2 simulation box (dashed lines) compared to the linear theory $\mathcal{P}(k)$ (solid lines). The times of the simulation snapshots correspond to the four panels of Fig.~\ref{fig:slices}. The grey dotted line marks $\mathcal{P} = 1$.}
\label{fig:pssim}
\end{figure}

In Figure \ref{fig:pssim}, we show the power spectra of the simulation particles for the four snapshots corresponding to the four panels in Fig.~\ref{fig:slices}. The panels show $\mathcal{P}(k)$ from the simulation as dashed lines and the linear theory $\mathcal{P}(k)$ as solid lines. For $a=0.10\rh{a}$, the smallest scales in the box are already nonlinear, as the power spectrum on these scales is higher than unity (shown by the grey dotted line). At $a=1.13\rh{a}$, the scales with $k > 10^8$ $h$/Mpc in the box have become nonlinear, which corresponds to the formation of the web of structures seen in the top right panel of \ref{fig:slices}.

At $a=11.27\rh{a}$, as the bottom left panel of Fig.~\ref{fig:slices} shows, erstwhile bound structures have mostly dissipated. This is shown in the fall-off of the simulation power spectrum in the bottom left panel of Fig.~\ref{fig:pssim}. The simulation power spectrum is much lower than the linear version for almost all scales. This suppression results from the free streaming of the $X$ particles after reheating. The bottom right panel of Fig.~\ref{fig:pssim} shows that the peak of the post-reheating power spectrum has moved to larger scales by $389.24\rh{a}$, confirming that the free-streaming cut-off has moved to larger scales with time. The slight rise in the simulation power spectrum at the highest $k$ in the bottom two panels of Fig.~\ref{fig:pssim} arises due to the Poisson noise of the discrete simulation particles. It is also apparent from Fig.~\ref{fig:pssim} that the free-streaming cut-off function is much shallower than the Gaussian cut-off used in Ref.~\cite{blanco19}, allowing significant power to persist more than a decade in $k$ beyond the scale at which suppression begins.

\section{Analyzing The Free-Streaming Cut-off}
\label{cut-off}

\begin{figure}[htb!]
\centering
\includegraphics[width=\textwidth]{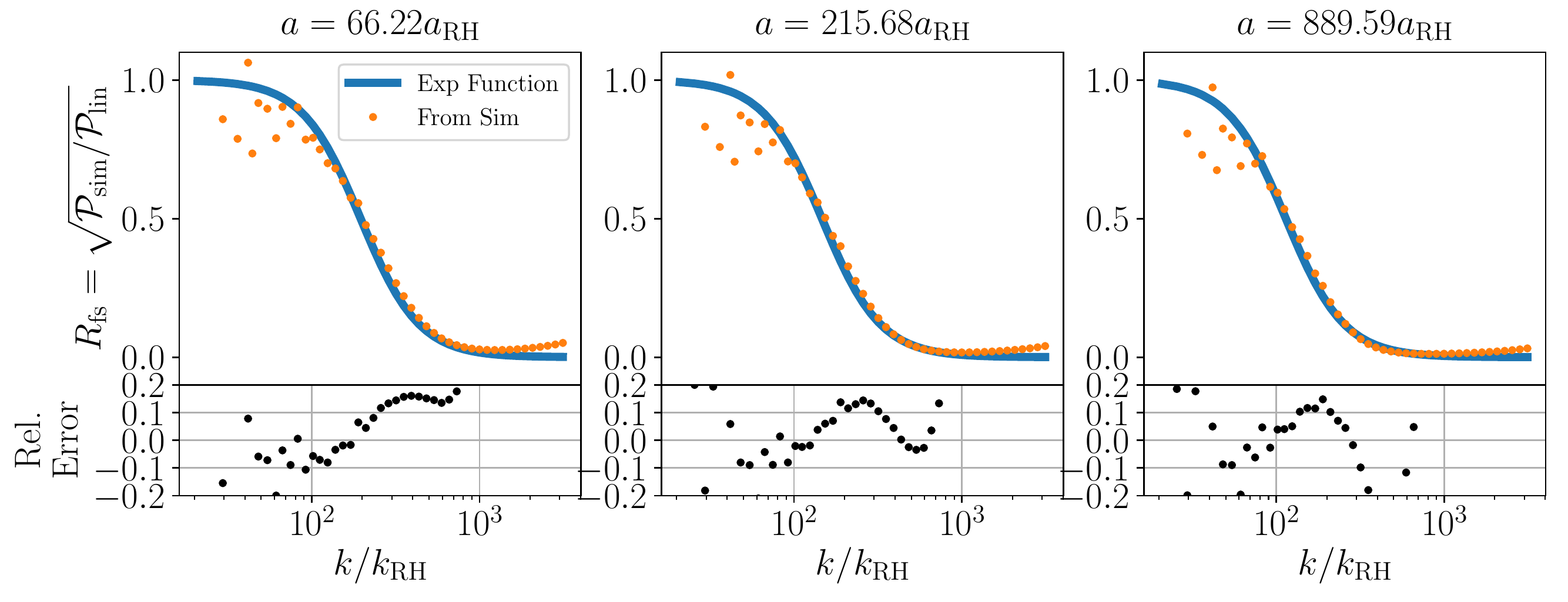}
\caption{$\rfs = \sqrt{\mathcal{P}_{\rm sim}/\mathcal{P}_{\rm lin}}$ from the RD1 simulation box, shown at different times after reheating. The blue curves show the function given by Eq.~(\ref{rfseq}) with $n=2.5$, while the bottom panels show the relative error between the $\rfs$ from simulations and the blue curves.
}
\label{fig:rfsfit}
\end{figure}

We quantify the free-streaming cut-off induced by gravitational heating as the ratio between the simulation and linear power spectrum as a function of time for each of our simulation runs, given by\footnote{Since $\mathcal{P}_{\rm sim}$ is the nonlinear power spectrum, the $\rfs$ inferred from Eq.~(\ref{simcut}) could be influenced by nonlinear effects instead of only free streaming. However, we find that this is not a significant concern. For example, in Fig.~\ref{fig:rfsfit}, $\mathcal{P}_{\rm sim}$ peaks at 0.4 in the first panel and 0.2 in the last, and the form of the cut-off is nevertheless essentially the same between them.} \begin{equation}\label{simcut}
    \rfs (k,a) \equiv \sqrt{ \frac{\mathcal{P}_{\rm sim} (k,a)}{\mathcal{P}_{\rm lin} (k,a)}},
\end{equation} for $a>10\rh{a}$. 
Here we evaluate the linear-theory power spectrum $\mathcal{P}_{\rm lin} (k,a)$ by using the Boltzmann solver described in G23 to obtain $\mathcal{P}_{\rm lin} (k,10\rh{a})$ and extrapolating it to later times using $\mathcal{P}_{\rm lin} (k,a)\propto [\ln(a/\rh{a})]^2$.

\begin{figure}[htb!]
\centering
\includegraphics[scale=0.85]{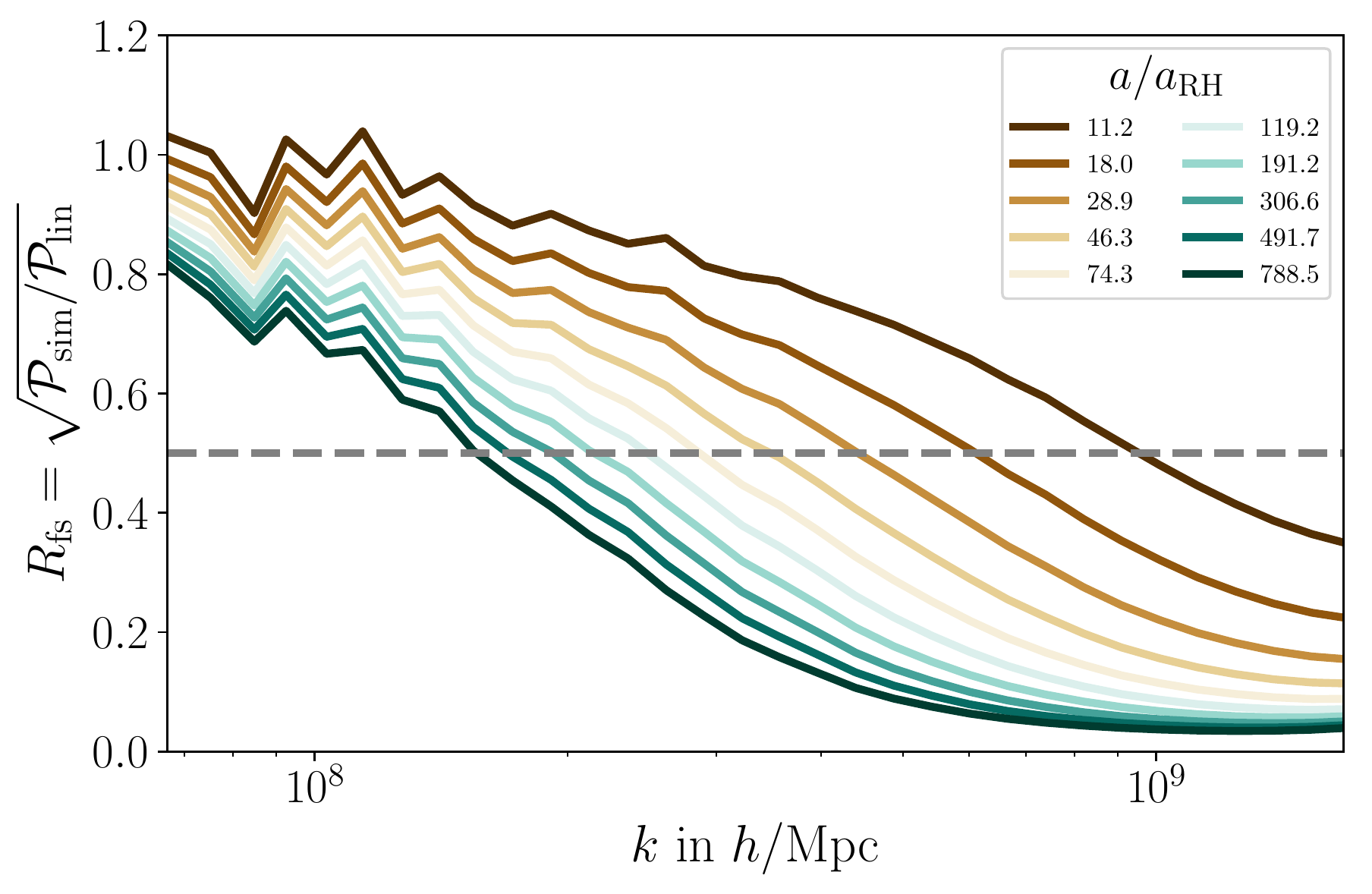}
\caption{Free-streaming cut-off $\rfs = \sqrt{\mathcal{P}_{\rm sim}/\mathcal{P}_{\rm lin}}$ from the RD5 simulation box, shown at different times after reheating. The grey dashed horizontal line indicates $\rfs = 0.5$. }
\label{fig:rfs}
\end{figure}

We analyzed snapshots from $11\rh{a}$ to $790 \rh{a}$ for each case. Figure \ref{fig:rfs} shows $\rfs$ at different times taken from the RD5 simulation box. As $a/\rh{a}$ increases, $\rfs(k,a)$ decreases and the wavenumber at which $\rfs = 0.5 $ (dashed line) moves to smaller $k$, indicating that the free-streaming cut-off to the power spectrum moves to larger scales.
The noise at low $k$ (also visible in Fig.~\ref{fig:pssim}) is due to the low number of modes sampled at scales close to the simulation box size and is not related to free streaming.

For each snapshot, $\rfs(k,a)$ was computed and fit to a function of the form \begin{equation} \label{rfseq}
    \frac{1}{1 + (k/\kfs)^{n}},
\end{equation} where $\kfs$ and $n$ were treated as free parameters. We found that $n \approx 2.5$ fits the fall-off in $\rfs$ for the range of our cases, so we subsequently fixed $n=2.5$ and fit this function to the simulation $\rfs$ while treating only $\kfs$ as a fitting parameter. Figure \ref{fig:rfsfit} shows the $\rfs$ obtained from the RD1 simulation (orange dots) at three different times. The blue curve in each panel shows the fit to Eq.~(\ref{rfseq}) with $n=2.5$. The bottom panels show the relative error between the simulation-based $\rfs$ values and the fit function. For the range of wavenumbers in which $\rfs$ drops, the error stays roughly within a 10\% bound, exceeding 10\% only when $\rfs \simeq 0.25$. As in Fig.~\ref{fig:rfs}, the noise at low $k$ is an artifact of the finite box size and is unrelated to free streaming.

\begin{figure}[htb!]
\centering
\includegraphics[width=\textwidth]{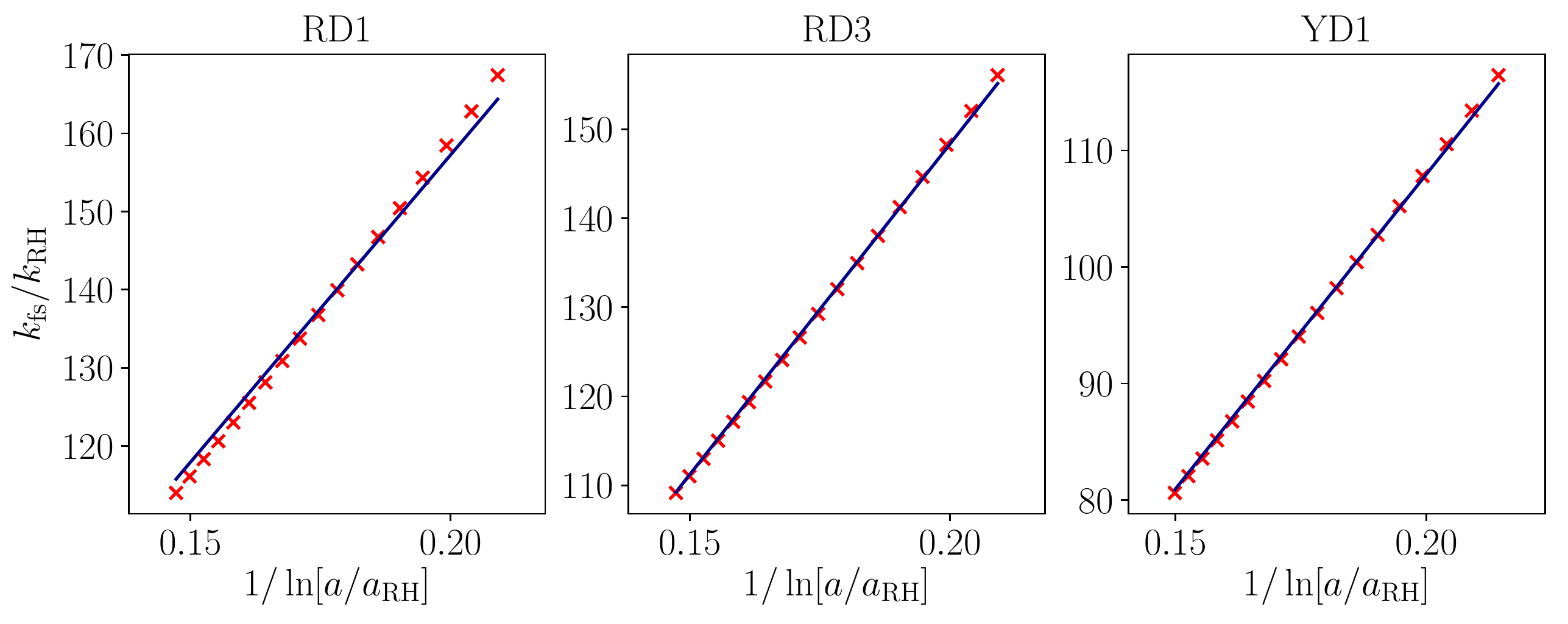}
\caption{The best fit values of $\kfs$ for the RD1, RD3 and YD1 simulation runs, with $[\ln (a/\rh{a})]^{-1}$ on the x-axis. The red crosses show the $\kfs$ values obtained by fitting the $\rfs$ from the simulations to Eq.~(\ref{rfseq}) with $n=2.5$. The dark blue lines show the best fit of these $\kfs$ values to a function proportional to $[\ln (a/\rh{a})]^{-1}$. }
\label{fig:kfsfit}
\end{figure}

This fitting procedure yields a value of $\kfs$ for each value of $a$. Figure \ref{fig:kfsfit} shows the values of $\kfs / \rh{k}$ obtained by this fitting for the RD1, RD3 and YD1 cases as red crosses, plotted against $[\ln (a/\rh{a})]^{-1}$. Guided by Eq.~(\ref{kfs}), we fit the array of $\kfs$ values obtained to the function \begin{equation}
    \frac{\kfs}{\rh{k}} = \alpha [\ln (a/\rh{a})]^{-1},
\end{equation} treating $\alpha$ as a fitting parameter. The dark blue lines in Fig.~\ref{fig:kfsfit} show the best fit lines using the above function. A total of 18 snapshots from $a \simeq 106 \rh{a}$ to $a \simeq 790 \rh{a}$ were used for this linear fit, in order to avoid the transients resulting from the decay of the $Y$ particle population and fit the free-streaming scale values only during radiation domination. For each of our nine simulation cases, we obtain the best fit value of $\alpha$, showing in Table \ref{tab:fitcoeff}.

\begin{table}[]
\centering
\begin{tabular}{|c|c|c|c|c|}
\hline
Case & $\sigma_{\rm lin,RH}$ (km/s) & Bound \% at $\rh{a}$ & $\alpha$ & $\sigma_{\rm lin,RH} \alpha/10^5$ (km/s)  \\ \hline \hline
RD1  & 851.26                    & 59.66      & 786.17   & 6.69                      \\ 
RD2  & 1738.77                   & 80.54      & 593.51   & 10.32                     \\ 
RD3  & 898.54                    & 36.02      & 741.59   & 6.66                      \\ 
YD1  & 2841.66                   & 86.72      & 539.78   & 15.34                     \\ 
YD2  & 7276.62                   & 97.23      & 554.65   & 40.36                     \\ 
YD3  & 1168.94                   & 32.84      & 611.03   & 7.14                      \\ 
YD4  & 1081.34                   & 27.46      & 630.61   & 6.82                      \\ 
RD4  & 385.80                    & 14.61      & 1650.05  & 6.37                      \\ 
RD5  & 377.40                    & 1.81       & 1636.21  & 6.18                      \\ \hline
\end{tabular}
\caption{The proportionality constant $\alpha$ relating $\kfs / \rh{k}$ and $[\ln (a/\rh{a})]^{-1}$ for the various simulation runs. Also listed are the linear velocity dispersion $\sigma_{\rm lin,RH }$ and the product of linear velocity dispersion and $\alpha$. Note that $\sigma_{\rm lin,RH }$ is calculated using Eq.~(\ref{siglin}) and converted to km/s from natural units.  }
\label{tab:fitcoeff}
\end{table}

\subsection{Relating the Cut-Off To The Linear Power Spectrum}

According to Eq.~(\ref{kfs}), the value of $\alpha$ is inversely proportional to the average velocity dispersion of the dark matter particles at reheating, denoted by $\rh{\sigma_v}$. To test this relation, we use the linear power spectrum evaluated at $\rh{a}$ to calculate the linear velocity dispersion for each of our nine cases. The power spectrum at $\rh{a}$ determines the linear velocity dispersion, which we take to be \begin{equation} \label{siglin}
    \sigma_{\rm lin,RH}\equiv \rh{k} \left( \int_0^\infty \frac{\diff k}{k} \frac{\mathcal{P}(k,\rh{a})}{k^2}\right)^{1/2}
\end{equation} based on the derivation given in Appendix \ref{linsigv} (we leave out the numerical factor for simplicity). Table \ref{tab:fitcoeff} shows the linear velocity dispersion for each case. Also shown is the product $\sigma_{\rm lin,RH} \alpha$. 

Two distinct classes can be identified from Table \ref{tab:fitcoeff}. For the six cases RD1, RD3, YD3, YD4, RD4 and RD5, the product $\sigma_{\rm lin,RH} \alpha$ has an average of $6.63 \times 10^5$ km/s, with the maximum deviation of the product from the average being 7.7\% (YD3). Within this error, for these cases, $ \alpha \propto 1/\sigma_{\rm lin,RH}$. For the other three cases, the product values are higher than those of the other six cases, and the product increases as the predicted bound fraction at reheating increases. 

The three cases with the high $\sigma_{\rm lin,RH} \alpha$ values are those with significant structure having formed at $\rh{a}$. 
Since Eq.~(\ref{siglin}) is valid in the linear regime, it is not expected to hold in these scenarios. As particles become bound into virialized structures, their velocities stop growing, so in general the correct nonlinear velocity dispersion is smaller than the linear extrapolation. The most massive halos have the largest virial velocities, so we expect them to dominate the nonlinear contributions to the velocity dispersion.

This motivates a truncated integral for the velocity dispersion, in which the integrand of Eq.~(\ref{siglin}) stops contributing for modes with $k>k_*$, where $\mathcal{P}(k_*,\rh{a})$ crosses some threshold value $\mathcal{C}$. We define the truncated linear velocity dispersion to be \begin{equation} \label{truncsigv}
    \sigma_{\rm t,RH} \equiv \rh{k} \left( \int_0^\infty \frac{\diff k}{k} \frac{\mathcal{P}_{\rm t}(k,\rh{a})}{k^2}\right)^{1/2},
\end{equation} where \begin{equation} \label{truncP}
    \mathcal{P}_{\rm t} \equiv \begin{cases}
        \mathcal{P}, &\  \mathcal{P}<\mathcal{C} \\
        \mathcal{C}^2 / \mathcal{P},  &\  \mathcal{P}>\mathcal{C}
    \end{cases} 
\end{equation} is the truncated linear power spectrum. The only motivation of this form is that it suppresses the contribution of modes with $\mathcal{P}\gg\mathcal{C}$ in a continuous way and fits the simulation results well, as we will see.

\begin{table}[]
\begin{tabular}{|c|c|c|c|c|c|}
\hline
Case & Bound \% at $\rh{a}$ & $\sigma_{\rm t,RH}$ (km/s) & $\alpha$ & $\sigma_{\rm t,RH} \alpha / 10^5$ (km/s) & \shortstack[c]{\% diff. from \\ avg. $\sigma_{\rm t,RH}\alpha$} \\ \hline \hline
RD1  & 59.66      & 851.26                     & 786.17                & 6.69                       & 0.04                                      \\ 
RD2  & 80.54      & 1130.70                    & 593.51                & 6.71                       & 0.31                                      \\ 
RD3  & 36.02      & 898.54                     & 741.59                & 6.66                       & 0.40                                      \\ 
YD1  & 86.72      & 1245.33                    & 539.78                & 6.72                       & 0.48                                      \\ 
YD2  & 97.23      & 1252.44                    & 554.65                & 6.95                       & 3.84                                      \\ 
YD3  & 32.84      & 1168.94                    & 611.03                & 7.14                       & 6.77                                      \\ 
YD4  & 27.46      & 1081.34                    & 630.61                & 6.82                       & 1.93                                      \\ 
RD4  & 14.61      & 385.80                     & 1650.05               & 6.37                       & 4.84                                      \\ 
RD5  & 1.81       & 377.40                     & 1636.21               & 6.18                       & 7.70                                      \\ \hline
\end{tabular}
\caption{The truncated linear velocity dispersion given by Eq.~(\ref{truncsigv}) with a threshold of $\mathcal{C} = 4.3$ (c.f. Table~\ref{tab:fitcoeff}). The second column from the right lists the product of $\sigma_{\rm t,RH}$ and $\alpha$. The rightmost column shows the percentage by which the product differs from the average product in the nine cases. The percentage difference is within 7.7\%, demonstrating that $\sigma_{\rm t,RH}$ is inversely proportional to $\alpha$. }
\label{tab:truncsig}
\end{table}

Various values of $\mathcal{C}$ were tried to obtain values of $\sigma_{\rm t,RH}$ that would bring the product $\sigma_{\rm t,RH} \alpha$ for the three high-bound-fraction cases closer to the products in the six other cases. We found $\mathcal{C} = 4.3$ to be an appropriate threshold. Table \ref{tab:truncsig} shows the truncated linear velocity dispersion and the products $\sigma_{\rm t,RH} \alpha$ for all nine cases with $\mathcal{C} = 4.3$. This truncation of the linear velocity dispersion does not affect the six cases in which the bound fraction at reheating is less than $80\%$, but the values of $\sigma_{\rm t,RH} \alpha$ are close to each other for all the nine cases, with a maximum deviation of 7.7\% from the mean value. 

This prescription of calculating $\sigma_{\rm t,RH}$ yields a velocity dispersion that is inversely proportional to $\alpha$ within 7.7\% error for our nine cases, covering the wide range of scenarios, with the bound percentages at $\rh{a}$ ranging from 1.81\% to 97.23\%. For the range of cases studied in this work, the free-streaming cut-off is given by \begin{equation} \label{kfsfit}
    \frac{\kfs (a)}{\rh{k}} = \frac{2.23}{\sigma_{\rm t,RH}} \left[ \ln \left( \frac{a}{\rh{a}} \right) \right]^{-1},
\end{equation} where the numerical factor is the average of the $\sigma_{\rm t,RH} \alpha$ values from Table \ref{tab:truncsig} (converted to units of $c$).

\begin{figure}[htb!]
\centering
\includegraphics[width=\textwidth]{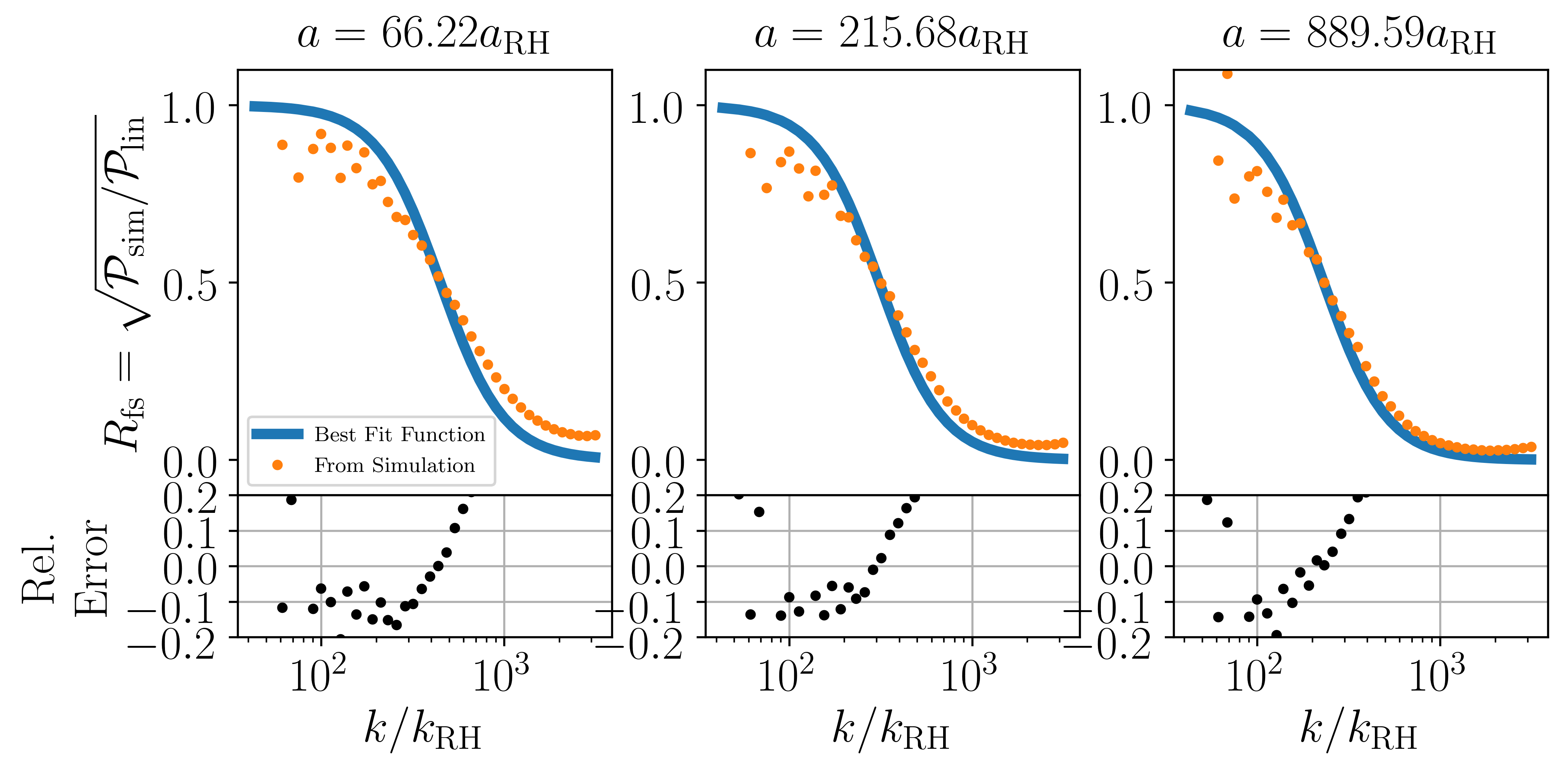}
\caption{$\rfs = \sqrt{\mathcal{P}_{\rm sim}/\mathcal{P}_{\rm lin}}$ from the RD4 simulation box, shown at different times after reheating. The blue curves show the function given by Eq.~(\ref{rfseq}) with $n=2.5$, while the bottom panels show the relative error between the $\rfs$ from simulations and the blue curves. }
\label{fig:rd4fits}
\end{figure}

For the RD4 and RD5 cases, since the amount of structure formed at $\rh{a}$ is not very high, the free-streaming fall-off in $\rfs$ is a little shallower than a $(k/\kfs)^{-2.5}$ power law. Figure \ref{fig:rd4fits} shows $\rfs$ from the RD4 simulations compared to the fitting form given by $(1 + (\kfs/\rh{k})^{2.5})^{-1}$. At $66.22\rh{a}$, the slopes of the fall-off from the simulation and the fitting function are noticeably different. At $889.59 \rh{a}$, the slopes are much closer to each other. The relative error between the $\rfs$ from the simulation and the best fit curve remains within 10\% for $\rfs \gtrsim 0.25$. Cases with such low amounts of bound structure at $\rh{a}$ present the boundaries down to which the $(1 + (\kfs/\rh{k})^{2.5})^{-1}$ fitting form is valid. For cases with lower bound percentages at $\rh{a}$, the fall-off in $\rfs$ is much shallower than $(k/\kfs)^{-2.5}$; we leave the detailed study of these cases to future work.

Finally, we remark that due to discreteness noise, it is difficult to resolve the precise form of the free-streaming cut-off for $k\gg\kfs$. This regime can nevertheless be important, since the linear power spectrum arising from an EMDE can grow so rapidly as a function of $k$.
Another possibility is to derive the form of the cut-off from the velocity distribution in the simulations. We use this approach in Appendix~\ref{velocity_distribution} to obtain an alternative expression for the cut-off function $\rfs$ and the free-streaming scale $\kfs$.
For the remainder of this article, however, we will continue to use the simple cut-off expression given by Eq.~(\ref{rfseq}) (with free-streaming scale given by Eq.~\ref{kfsfit}).

\section{Structure Formation After Reheating}
\label{latestructure}

With the shape and scale of the free-streaming cut-off ascertained, we can obtain the matter power spectrum in the post-EMDE era in scenarios with gravitational heating. The power spectrum in turn determines the microhalo distribution that arises in connection with the scenario.

We first make a technical note. The expression $\lfs \propto \ln(a/\rh{a})$ is derived under the assumption of radiation domination.
During the transition to late matter domination,
\begin{align}\label{lfsMR}
    \lfs
    &\propto \int_{\rh{a}}^a
    \frac{\diff a^\prime}{a^{\prime} \sqrt{\aeq+a^\prime}}
    \nonumber\\
    &\propto \ln\left[\left(\frac{1+\sqrt{1+\rh{a}/\aeq}}{1+\sqrt{1+a/\aeq}}\right)^2\frac{a}{\rh{a}}\right]
    \simeq \ln\left[\left(\frac{2}{1+\sqrt{1+a/\aeq}}\right)^2\frac{a}{\rh{a}}\right]
\end{align}
(compare Eq.~\ref{lfs}), where we assume $\aeq\gg\rh{a}$ and again neglect changes to the radiation content.
This expression is approximately $\ln(a/\rh{a})$ when $a\ll\aeq$ but asymptotes to $\ln(4\aeq/\rh{a})$ in the $a\gg\aeq$ limit.
We therefore evaluate the free-streaming cutoff arising from gravitational heating by replacing $\ln(a/\rh{a})$ with the logarithm in Eq.~(\ref{lfsMR}) in the expression for the cut-off scale, Eq.~(\ref{kfsfit}).

\subsection{The linear matter power spectrum}

\begin{figure}[htb!]
\centering
\includegraphics[width=\textwidth]{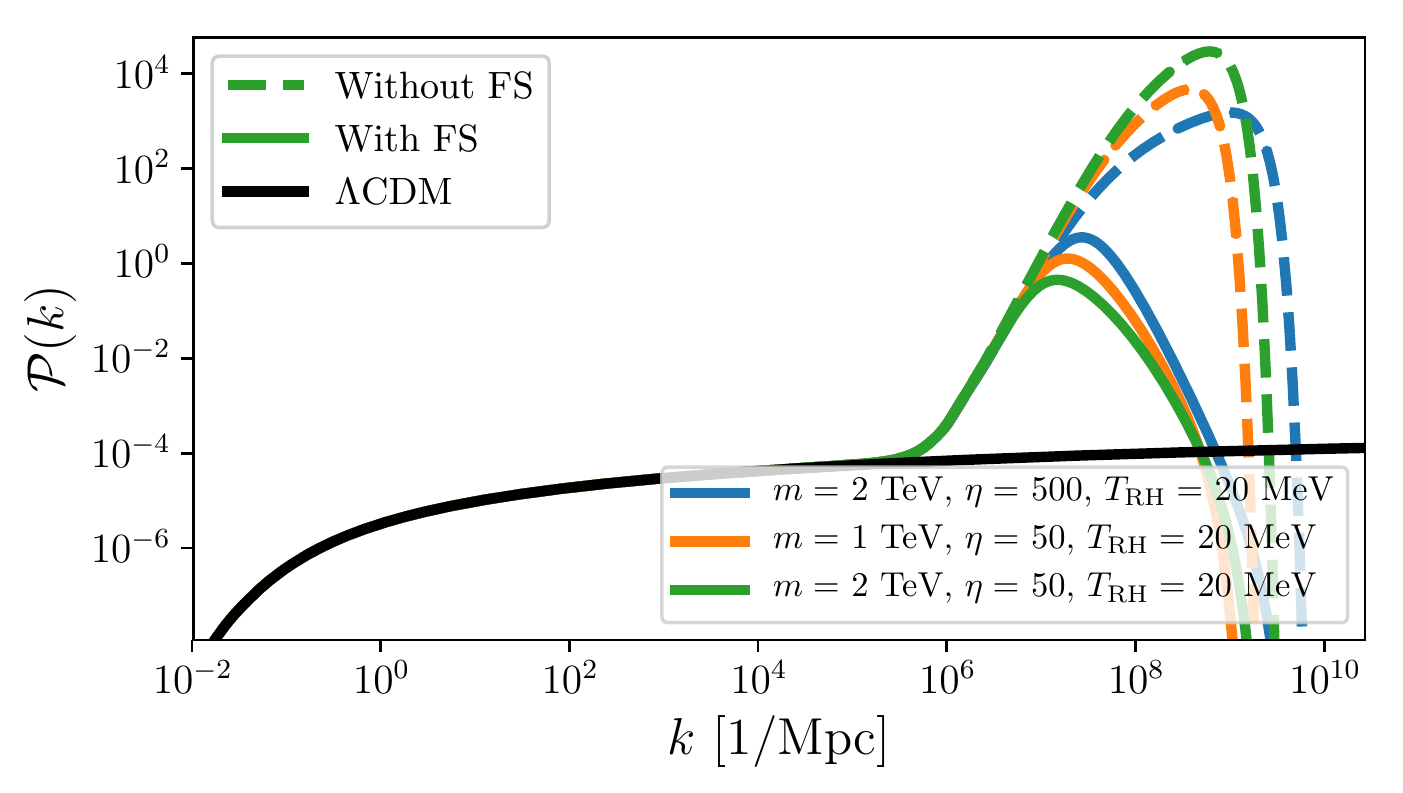}
\caption{The $\Lambda$CDM power spectrum (black) along with three EMDE-enhanced power spectra (dashed) and the same power spectra with the free-streaming cut-off (solid), all evaluated at $a = \aeq$.
}
\label{fig:aeq_3ps}
\end{figure}

Figure \ref{fig:aeq_3ps} shows the $\Lambda$CDM power spectrum $\mathcal{P}$ at $\aeq$ (black solid curve). We first evaluated the linear power spectrum using \textsc{CAMB} at $z=500$ and translated it back in time to $\aeq$ via the growth factor derived in Appendix \ref{growthfac}. This growth factor ignores the scale-dependent evolution of modes that enter the horizon after reheating, but the amplitudes of these modes are not affected by the EMDE, so they are largely irrelevant when evaluating how the EMDE influences structure formation.
The dashed colored curves show the power spectra without the effects of free-streaming for three EMDE cosmologies, computed by multiplying the $\Lambda$CDM power spectrum by the transfer functions from G23. The orange and green dashed curves are different only in the value of the $Y$ particle mass $m$. The shape of the power spectrum peak is similar for these two cases, since the value of $\eta$ is the same. The case with the lower $m$ (orange) has a larger cut-off scale from the pressure of the $Y$ particles compared to the green power spectrum. In contrast, the blue dashed curve is associated with the same $m$ and $\rh{T}$ as the green dashed curve, but its value of $\eta$ is much higher. This difference leads to a much shallower peak for the blue dashed curve.

The solid curves show the power spectra for the same EMDE cases with the free-streaming cut-off included by multiplying the dashed curves by the cut-off function $\rfs^2$ from Eq.~(\ref{rfseq}) with cut-off scale $\kfs$ given by Eq.~(\ref{kfsfit}) (modified as described after Eq.~\ref{lfsMR}) at $a=\aeq$. In all three cases, the power spectrum peaks move to much larger scales and smaller amplitudes due to the effects of gravitational heating. Nonetheless, a significant portion of the EMDE enhancement is retained in all cases. For example, the case with the green curves has a 65\% predicted bound fraction at reheating, and yet the power spectrum peak after including free-streaming effects is roughly four orders of magnitude larger than the $\Lambda$CDM counterpart at the same scale.

It is noteworthy that with gravitational heating accounted for, all three EMDE power spectra in Fig.~\ref{fig:aeq_3ps} peak around $\mathcal{P}\sim 1$ at matter-radiation equality. In fact, this is generally expected as long as the dimensionless linear power spectrum during the EMDE scales more steeply than $\mathcal{P}\sim k^2$, because then the velocity dispersion integral in Eq.~(\ref{siglin}) is dominated by high-$k$ contributions. That is, for such power spectra, modes that are just becoming nonlinear at reheating (as opposed to modes that are still linear) contribute the most to the velocity dispersion. This leads to $\sigma_{\rm t,RH}\sim \rh{k}/k_*$, where $k_*$ is the nonlinear scale (where $\mathcal{P}\sim 1$) at reheating. Equation (\ref{kfsfit}) then implies that $\kfs\sim k_*/\ln(a/\rh{a})$. Shortly after reheating, the power spectrum thus peaks at $\mathcal{P}\sim 1$ (see also Fig.~\ref{fig:pssim}), since $\kfs$ is close to the nonlinear scale $k_*$. Over time, $\kfs$ shifts to logarithmically larger scales, and the power spectrum also grows logarithmically (per Eq.~\ref{growth_RD}). The former effect tends to win, causing the peak in the power spectrum to slightly drop over time (as is evident in Fig.~\ref{fig:pssim}). For example, if $\mathcal{P}(k) = (k/k_*)^n$ at reheating, then the combination of linear growth and continued free streaming results in the power spectrum peaking at a value around $\mathcal{P}\sim [\ln(a/\rh{a})]^2(\kfs/k_*)^n\sim [\ln(a/\rh{a})]^{2-n}$ at later times during radiation domination, which drops over time as long as $n>2$. But since this drop is only logarithmic, we expect that the power spectrum still peaks at some value $\mathcal{P}\sim 1$ at the onset of late matter domination.

\begin{figure}[htb!]
\centering
\includegraphics[width=\textwidth]{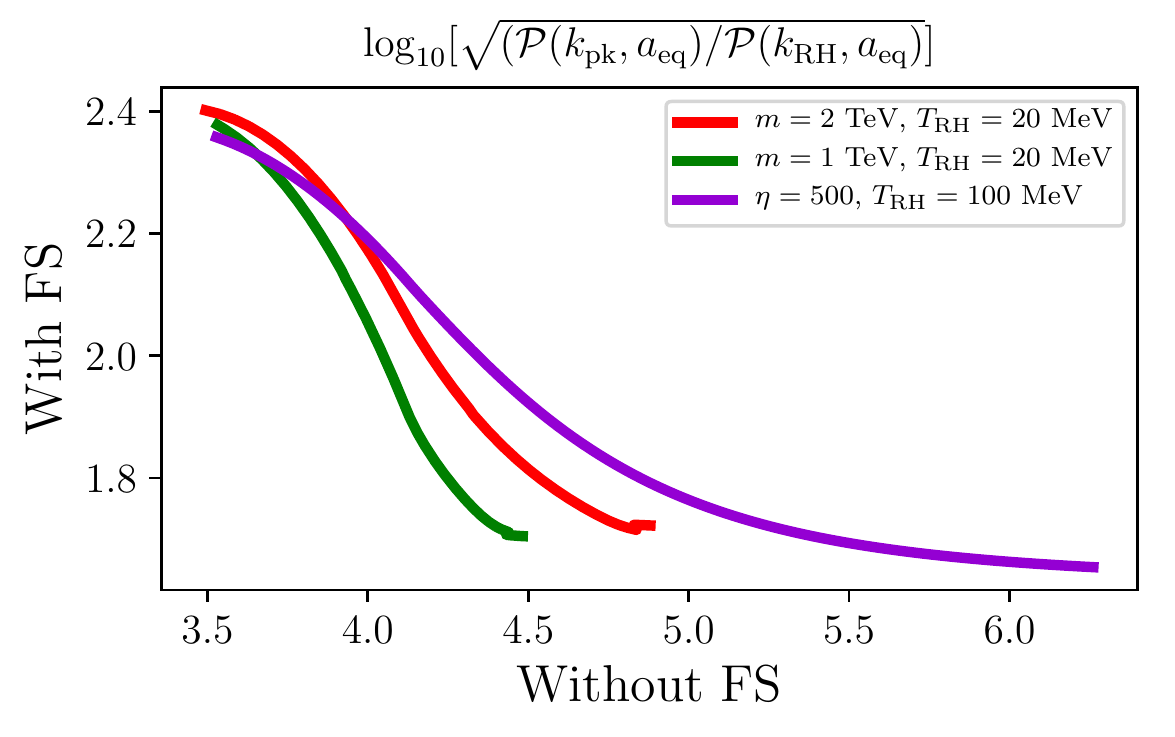}
\caption{The logarithm of $\sqrt{\mathcal{P}(k_{\rm pk},\aeq) / \mathcal{P}(\rh{k},\aeq) }$, which shows the peak enhancement to the power spectrum. The x-axis shows the peak with the effects of gravitational heating ignored, while the y-axis shows the peak with the free-streaming cut-off included for the same EMDE cosmology. The red and green curves are obtained by varying $\eta$ for different values of $m$, while the purple curve corresponds to varying $m$ with fixed $\eta$.
}
\label{fig:peaks}
\end{figure}

Nevertheless, Fig.~\ref{fig:aeq_3ps} shows a tendency where if the power spectrum without the gravitational heating suppression (dashed curves) peaks at a higher value, the peak with the suppression accounted for (solid curves) is lower. Figure~\ref{fig:peaks} demonstrates this trend more broadly. Here we plot the logarithm of $\sqrt{\mathcal{P}(k_{\rm pk},\aeq) / \mathcal{P}(\rh{k},\aeq) }$, the factor by which density variations at the scale of the peak in the power spectrum exceed density variations at $\rh{k}$ (which are not affected by the EMDE). This ratio quantifies the degree to which the EMDE boosts the power spectrum, and we plot its value with gravitational heating accounted for as a function of its value with gravitational heating neglected. Generally, as the unsuppressed value of $\sqrt{\mathcal{P}(k_{\rm pk},\aeq) / \mathcal{P}(\rh{k},\aeq) }$ rises from $\sim 10^{3.5}$, its gravitational heating-suppressed value drops from $\sim 10^{2.4}$.

More precisely, each of the three curves in Fig.~\ref{fig:peaks} represents a family of EMDE models where one parameter out of $m$, $\eta$ and $\rh{T}$ is varied while the others are held fixed. 
The red and green curves are obtained by varying $\eta$; as indicated by the comparison between the blue and green dashed curves in Fig.~\ref{fig:aeq_3ps}, smaller values of $\eta$ raise the peak value of $\mathcal{P}$ while shifting it to larger scales. Both of these effects raise the velocity dispersion during the EMDE, since it involves an integral over $k^{-2}\mathcal{P}(k)$ (see Eq.~\ref{siglin}), which explains why the gravitational heating-suppressed value of $\sqrt{\mathcal{P}(k_{\rm pk},\aeq) / \mathcal{P}(\rh{k},\aeq) }$ in Fig.~\ref{fig:peaks} drops sharply as a function of its unsuppressed value.
In contrast, the purple curve is obtained by varying the $Y$ particle mass $m$; as illustrated by the orange and green dashed curves in Fig.~\ref{fig:aeq_3ps}, when $\eta$ is large, increasing $m$ raises the peak value of $\mathcal{P}$ while shifting it to smaller (not larger) scales. These two changes to the power spectrum influence the velocity dispersion during the EMDE oppositely -- increasing $\mathcal{P}$ raises it while moving the peak to smaller scales lowers it -- which is why the purple curve in Fig.~\ref{fig:peaks} drops more shallowly.

Figure~\ref{fig:peaks} suggests that the free-streaming-suppressed value of $\sqrt{\mathcal{P}(k_{\rm pk},\aeq) / \mathcal{P}(\rh{k},\aeq) }$ does not drop below a factor of $\sim 10^{1.6}$. The presence of such a limit is explained because it corresponds to the scenario where the unsuppressed linear power spectrum at reheating is $\mathcal{P}(k)\simeq 0.15 A_s (k/\rh{k})^4$ down to arbitrarily small scales (high $k$; compare Fig.~\ref{fig:ps}). Here $A_s$ is the primordial spectral amplitude; we neglect the spectral tilt for simplicity. Since the truncated linear velocity dispersion relevant to gravitational heating, given by Eq.~(\ref{truncsigv}), involves an integral only over scales that are linear or mildly nonlinear ($\mathcal{P}\lesssim 1$), it converges to a finite value that evaluates to $\sigma_\mathrm{t,RH}\simeq 0.73 A_s^{1/4}\simeq 0.005$ (1500~km/s), assuming $A_s=2.2\times 10^{-9}$.
The free-streaming scale at matter-radiation equality is then $\kfs\sim \sigma_\mathrm{t,RH}^{-1}[\ln(\aeq/\rh{a})]^{-1}\rh{k}$, and thus $\sqrt{\mathcal{P}(k_{\rm pk},\aeq) / \mathcal{P}(\rh{k},\aeq) }\sim (\kfs/\rh{k})^2\sim \sigma_\mathrm{t,RH}^{-2}[\ln(\aeq/\rh{a})]^{-2}$.
Since $\sigma_\mathrm{t,RH}\simeq 0.005$ and $\ln(\aeq/\rh{a})\sim\mathcal{O}(10)$, small-scale density variations remain strongly enhanced even in this limiting scenario.

\subsection{Dark matter halo formation}

To illustrate how the gravitational heating cut-off affects halo formation, we consider the halo mass function $\diff n / \diff \ln M$, the number density of bound halos in logarithmic bins of mass, at different times around the epoch of matter-radiation equality. According to Press-Schechter theory \cite{press-schechter},
\begin{equation}
    \frac{\diff n}{\diff \ln M} =
    \sqrt{\frac{2}{\pi}}
    \frac{\bar{\rho}_X}{M}
    \left| \frac{\diff \ln \sigma}{\diff \ln M} \right|
    \frac{\delta_c}{\sigma(M)}
    \exp \left[- \frac{\delta_c^2}{2 \sigma^2(M)} \right],
\end{equation}
where $\sigma(M)$ is the rms density contrast smoothed on the mass scale $M$ and $\bar{\rho}_X$ is the comoving density of dark matter.
Following common practice for cosmologies with a small-scale cut-off to the power spectrum, we use a sharp $k$-space smoothing filter and evaluate
$\sigma^2(M) = \int_0^{k_M} \mathcal{P}(k) \diff k/k$
with the smoothing wavenumber $k_M$ defined by $M \equiv (4\pi/3)\bar{\rho}_X (2.5/k_M)^3$; the factor of 2.5 is motivated in Ref.~\cite{benson12} by a match to simulations, since there is no natural relationship between length and mass scales for this smoothing filter.
Since we will focus on times close to matter-radiation equality, we use the time-dependent fitting function given in Ref.~\cite{blanco19} for the spherical collapse threshold, $\delta_c$, which is valid for mixed matter-radiation domination.

\begin{figure}[htb!]
\centering
\includegraphics[width=\textwidth]{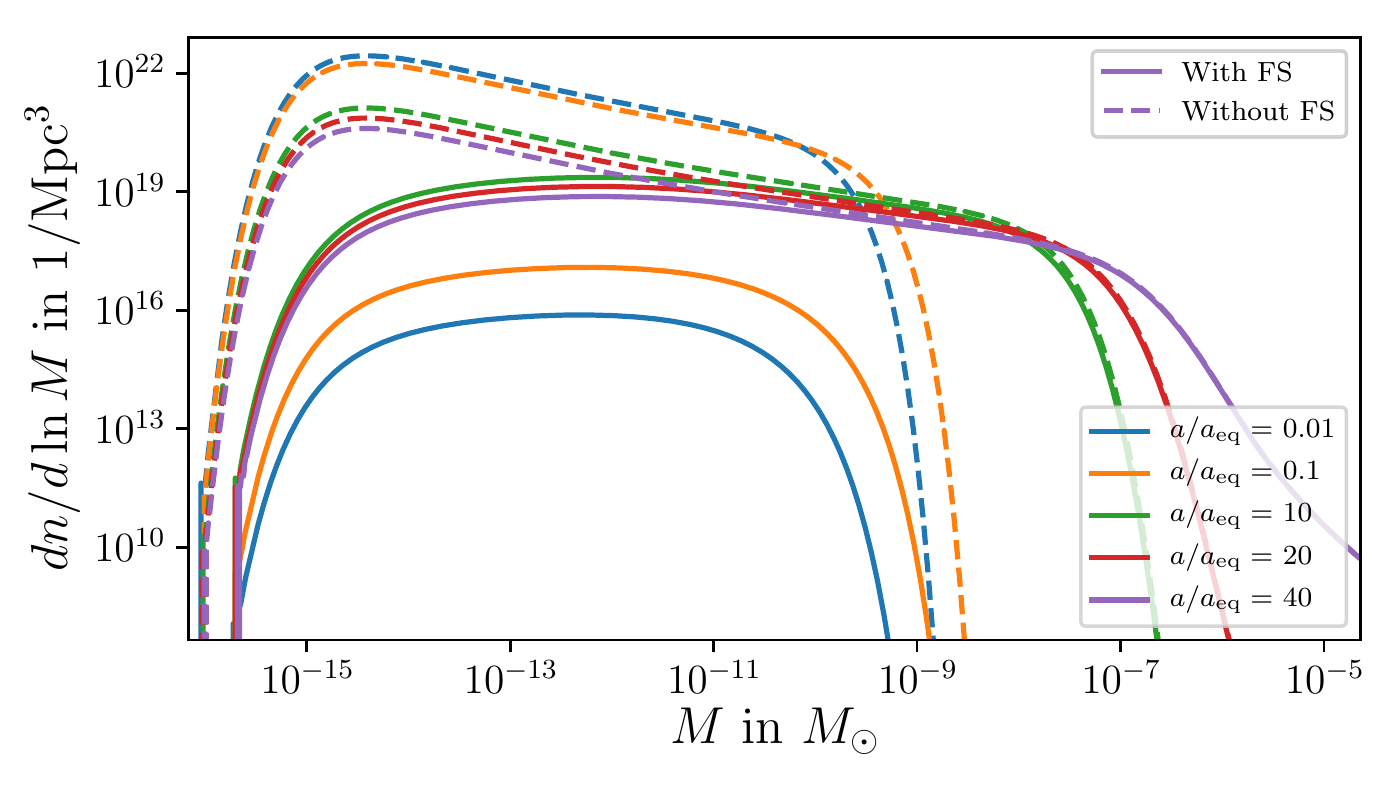}
\caption{The halo mass function, or differential number density of halos per logarithmic mass interval, as a function of mass at different times around matter-radiation equality for an EMDE scenario with $m=2$ TeV, $\eta = 50$, $\rh{T} = 20$ MeV. The solid curves show the mass function evaluated with the free-streaming cut-off, while the dashed curves show the version calculated without the cut-off, with the EMDE enhancement intact. }
\label{fig:hmf1}
\end{figure}

Figure \ref{fig:hmf1} shows the mass function for the case $m=2$ TeV, $\eta = 50$ and $\rh{T} = 20$ MeV, the same cosmology as the green curves in Figure~\ref{fig:aeq_3ps}. The solid lines show the mass function with the gravitational heating cut-off imposed on the power spectrum using the results of Sec~\ref{cut-off}. The dashed lines show the version without the cut-off, with the EMDE enhancement intact. The dashed lines are significantly higher than their solid counterparts for low masses, showing the formation of a significantly higher number of halos at small scales. This phenomenon is strongly inhibited by the effects of heating. 
The solid lines show a high density of bound halos of sub-Earth mass, even as early as $0.1\aeq$. For $a<\aeq$, the mass function shown by the solid lines grows as more small-scale halos form. For $a>\aeq$, the same mass function decreases with time for $M \lesssim 10^{-8} M_{\odot}$ and increases for larger masses as smaller halos merge to make larger ones. For comparison, the mass function at the scales shown in Figure~\ref{fig:hmf1} is lower than $10^{-150} \rm{ Mpc }^{-3}$ at $\aeq$ if a $\Lambda$CDM power spectrum with a Gaussian cut-off at $10^6$ Mpc$^{-1}$ is used. The cut-off wavenumber imposed by gravitational heating is $\kfs \approx 40 \rh{k}$ for this EMDE scenario, which shows that even with a small portion of the EMDE-induced bump persisting, the number of small-scale halos is significantly higher than in cases without an EMDE.

\begin{figure}[htb!]
\centering
\includegraphics[width=\textwidth]{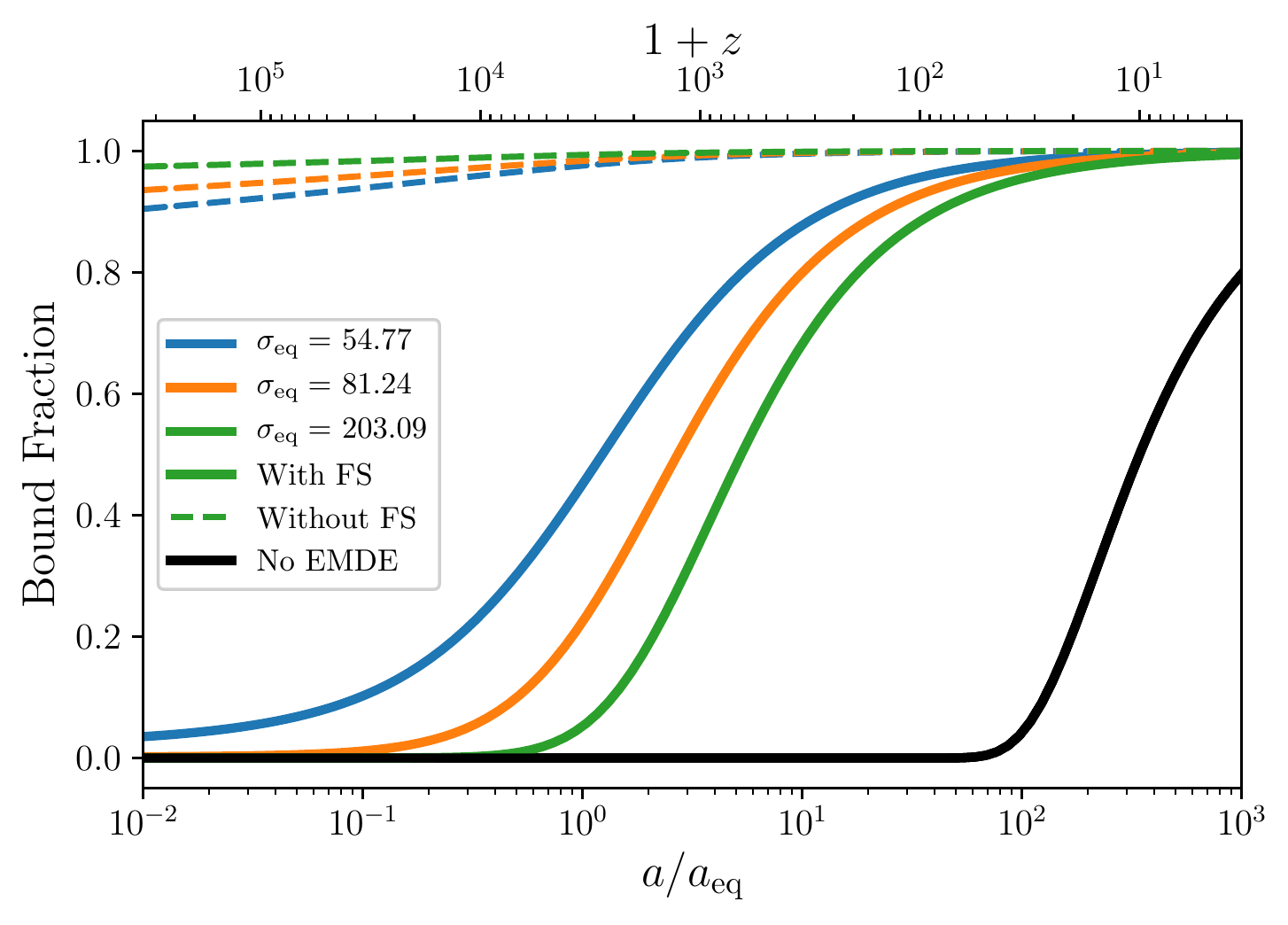}
\caption{
The bound fraction calculated using Press-Schechter theory at different times after matter-radiation equality for three EMDE scenarios, which are the same ones considered in Fig.~\ref{fig:aeq_3ps}. The amount of small-scale enhancement to the power spectrum due to the EMDE is quantified by $\seq$. The dashed lines show the bound fractions if no free-streaming cut-off is imposed on the power spectrum after the EMDE, while the solid lines show the bound fraction with the cut-off. For comparison, we also show a standard $\Lambda$CDM scenario (black curve), where we impose a Gaussian cut-off at $10^6$ Mpc$^{-1}$ (corresponding to the smallest microhalos being roughly earth-mass; e.g. Ref.~\cite{2004MNRAS.353L..23G}), although the result is only very weakly sensitive to this choice.
}
\label{fig:bfevol}
\end{figure}

In cases with higher enhancement on small scales due to the EMDE, more structure forms in the EMDE and the free-streaming scale at matter-radiation equality is larger. Therefore, the higher the EMDE enhancement, the greater the suppression of structure formation after the EMDE due to gravitational heating. Figure~\ref{fig:bfevol} illustrates this point. The fraction of dark matter in bound halos (Eq.~\ref{fbound}) is plotted versus $a/\aeq$ for the three cases in Fig.~\ref{fig:aeq_3ps}, which we now label using $\seq$, the value that the rms linear density fluctuation would reach at $a=\aeq$ if gravitational heating were neglected. That is,
\begin{equation}\label{seq}
    \seq \equiv \left[\int \mathcal{P}_{\rm emde}(k,\aeq) \diff \ln k\right]^{1/2},
\end{equation}
where $\mathcal{P}_{\rm emde}$ is the power spectrum without the free-streaming cut-off (corresponding to the dashed curves in Fig.~\ref{fig:aeq_3ps}). $\seq$ thus quantifies the extent of the enhanced perturbation growth during an EMDE.

The dashed curves show the predicted bound fraction for each case with gravitational heating neglected. 
The bound fraction predicted under this assumption at any given time is always larger with higher $\seq$, and in all three cases nearly all of the matter is predicted to be bound in halos by the time of matter-radiation equality.
The solid curves show the same cases with the free-streaming cut-off from gravitational heating now imposed. For higher $\seq$, the bound fraction shown by the solid curves at a given time is lower. This inversion shows the effect of gravitational heating: more structure formation during the EMDE (higher $\seq$) leads to a larger free-streaming scale and hence more suppression of structure formation after the EMDE. For comparison, the solid black line shows the predicted bound fraction without the EMDE; here a $\Lambda$CDM power spectrum is assumed with a Gaussian cut-off at $10^6$ Mpc$^{-1}$, which corresponds to the smallest microhalos being earth-mass (see \cite{2004MNRAS.353L..23G} for an example). Even with the gravitational heating-induced suppression, the bound fraction is much higher at $a \approx 10\aeq$ in cases with an EMDE.

\begin{figure}[htb!]
\centering
\includegraphics[width=\textwidth]{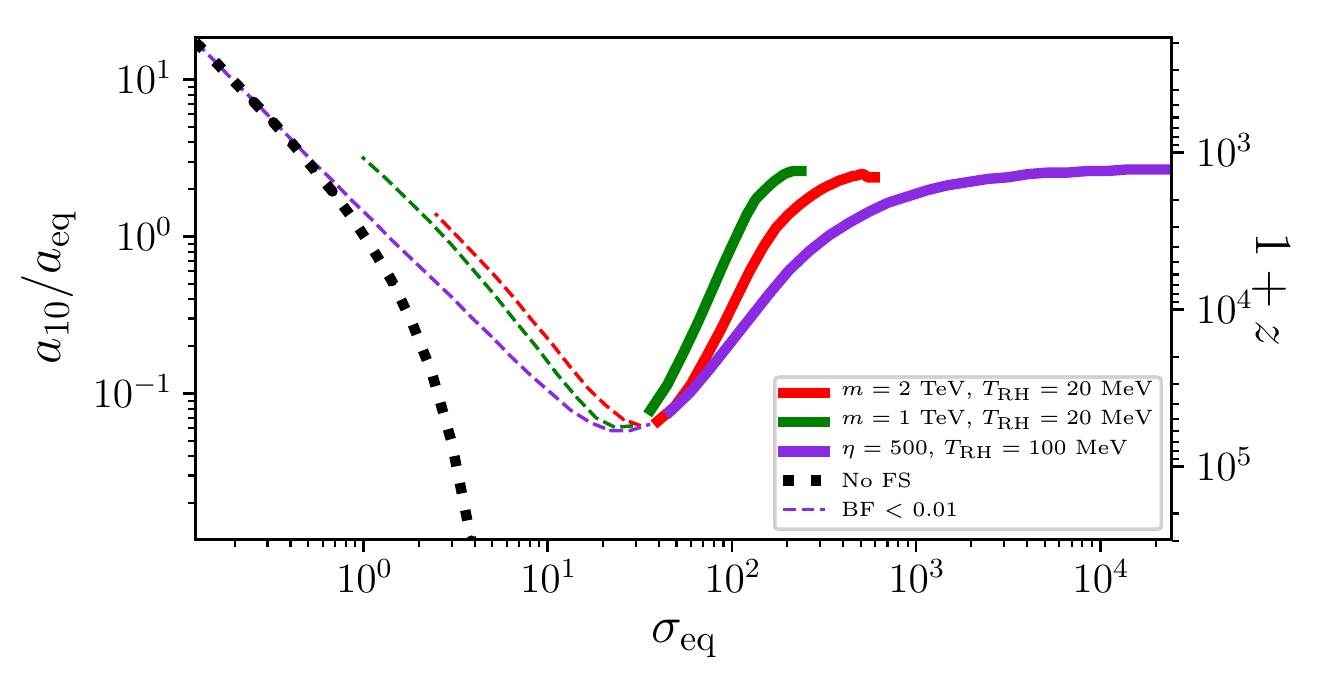}
\caption{
The scale factor at which 10\% of the dark matter is predicted to be bound in halos according to Press-Schechter theory, plotted against $\seq$ (Eq.~\ref{seq}), which quantifies the extent to which the small-scale power spectrum is enhanced during an EMDE. The curves are obtained by varying $\eta$ (red and green) and $m$ (purple) to vary $\seq$ by varying the shape and peak of the EMDE-enhanced power spectrum. The dashed curves show cases in which the bound fraction at $\rh{a}$ $< 0.01$, the regime in which our free-streaming cut-off prescription is untested. The dotted line shows $a_{10}$ for the same EMDE cases without the free-streaming cut-off, which only depends on $\seq$ regardless of the power spectrum shape.
}
\label{fig:a10}
\end{figure}

Figure \ref{fig:a10} shows, as a function of $\seq$, the time $a_{10}$ at which 10\% of the dark matter is predicted to be bound in halos after matter-radiation equality. By Eq.~(\ref{fbound}), this happens when the rms linear density contrast is $\sigma=0.608\dc$ ($\sigma=1.025$ during matter domination) when gravitational heating is accounted for.
The three colors here correspond to the same three families of EMDE model considered in Fig.~\ref{fig:peaks}, each obtained by changing one of $m$, $\eta$ and $\rh{T}$ while keeping the others fixed.
The solid curves account for the free-streaming cut-off imposed by gravitational heating and mark the scenarios where at least 1 percent of the matter is bound into halos by reheating (corresponding to $\seq\gtrsim 40$). In this regime, increasing $\seq$ raises the bound fraction at reheating, leading to significant gravitational heating and the consequent erasure of some of the EMDE-induced boost to the power spectrum. This erasure causes a reduction in the peak amplitude of the power spectrum (as we showed previously in Fig.~\ref{fig:peaks}), and this results in later formation of nonlinear structure after reheating.
The same trends are apparent in Fig.~\ref{fig:a10} as in Fig.~\ref{fig:peaks}; varying $\eta$ causes $a_{10}$ to change more mildly as a function of $\seq$, compared to varying $m$, due to how the scale of the peak in $\mathcal{P}(k)$ shifts as a result of these changes. As discussed above, the EMDE-induced enhancement to small-scale density variations remains large even in scenarios with an arbitrarily large degree of structure formation during the EMDE. That outcome is reflected in Fig.~\ref{fig:a10} in how $a_{10}$ remains below about $3\aeq$ (redshift 1000) in all cases.

The dotted curve in Fig.~\ref{fig:a10} shows how $a_{10}$ varies with $\seq$ when gravitational heating is neglected. Under this assumption, $\seq$ completely determines the unsmoothed rms density contrast and hence the bound fraction during the late matter era (see Eq.~\ref{fbound}), so there is only one curve.
Higher $\seq$ leads to earlier halo formation, and when $\seq\gtrsim 1$, a significant halo population forms during the radiation era,\footnote{But deviations from spherical symmetry are particularly important for halo formation during radiation domination \cite{2013JCAP...11..059B,2023MNRAS.520.4370D,2023PhRvD.107h3505D}, so our use of the spherical collapse threshold may yield inaccurate results in this regime.} but only as long as no significant nonlinear structures arise during the EMDE.
Further work is needed to understand how the dotted curve (with no gravitational heating) transitions into the solid curves (with gravitational heating) in the $1\lesssim\seq\lesssim 40$ regime.
As an illustration, the thin dashed curves in Fig.~\ref{fig:a10} use our gravitational heating prescription to extend the solid curves to include cases with less than 1\% of the matter bound in halos at reheating. However, the prescription is untested in this regime, and indeed it is unclear whether the underlying principles should continue to hold. If density variations remain linear during the EMDE, then even though particle velocities are boosted, they remain associated with coherent fluid motions rather than the velocity dispersion that is required to produce a free-streaming cut-off.
Nevertheless, the qualitative outcome should hold that higher $\seq$ results in earlier halo formation until $\seq\sim\mathcal{O}(10)$, beyond which point the trend should reverse.

\section{Summary And Discussion}
\label{summary}

In cosmologies with an early matter-dominated era (EMDE) prior to Big Bang Nucleosynthesis, the small-scale power spectrum is enhanced due to the linear growth of subhorizon matter perturbations with scale factor \cite{ae15}. This enhancement causes the formation of dense microhalos much earlier than bound structures form in non-EMDE scenarios \cite{emde2,emde1,ae15}.
These new sub-Earth-mass structures boost the dark matter annihilation signal \cite{ae15,blanco19,sten_gr} and can be detected gravitationally via caustic microlensing \cite{caustic1,caustic2,caustic3,obs_blinov} and pulsar timing arrays \cite{pta1,obs_blinov,pta_sten}.
However, if perturbations are sufficiently boosted, microhalos can form during the EMDE itself. They would dissipate when the EMDE ends, releasing their
dark matter particles at boosted speeds in random directions \cite{blanco19}. This gravitational heating imprints a free-streaming cut-off on the matter power spectrum after the EMDE, which suppresses some of the EMDE-induced enhancement to the power spectrum. Ascertaining the nature of this free-streaming cut-off is thus essential for understanding the halo populations that arise from these EMDE scenarios and hence their observational constraints and prospects.

To analyze the shape and scale of the free-streaming cut-off, we employed a modified version of \textsc{GADGET-2} to run the first suite of $N$-body simulations of the process of microhalo formation during the EMDE and their evaporation at the end of the EMDE. We found the formation of a rich web of small-scale structure during the EMDE, which gets wiped out as structures evaporate when the particle driving the EMDE decays and radiation domination starts.
By analyzing the power spectra of our simulations during the subsequent radiation era, we found a free-streaming cut-off that evolved to larger scales as time elapsed.
The various simulation runs adopted different EMDE model parameters in order to cover a range of scenarios, including variations in the level of structure formation during the EMDE.

We found that free streaming due to gravitational heating suppresses the matter power spectrum $\mathcal{P}(k)$ during the subsequent radiation epoch by a factor that is fit well by the function $[1 + (k/\kfs)^{2.5}]^{-2}$ (or an alternative form given in Appendix~\ref{velocity_distribution}).
The free-streaming wavenumber $\kfs$ can be straightforwardly calculated by means of Eq.~(\ref{kfsfit}); it is tightly connected to an expression given by Eq.~(\ref{truncsigv}) for the velocity dispersion during the EMDE.
Given the linear power spectrum due to an EMDE (expressions for which are provided in Ref.~\cite{hg23}), these results enable easy analytic computation of the impact of gravitational heating.

Finally, we used our results to explore how gravitational heating impacts structure formation during the late matter epoch. Free streaming erases a large portion of the EMDE-induced boost to the linear matter power spectrum; its peak value at the time of matter-radiation equality, $\aeq$, can be 4-5 orders of magnitude smaller than its peak value at the end of the EMDE. Despite this, sufficient power typically persists to form a significant microhalo population around $\aeq$.
In cases where $\sim 1$ percent of the matter is bound in halos by the end of the EMDE, about 10 percent of the dark matter is expected to be in microhalos already as early as $a=0.1\aeq$. Without an EMDE, the formation of a similar amount of bound structure would not occur until $a\sim 100\aeq$ ($z\sim 30$).
As the level of structure formation during the EMDE is raised, the free-streaming scale due to gravitational heating also grows, which results in greater erasure of the EMDE-induced boost to the power spectrum and hence later microhalo formation times after the EMDE. However, even in cases where most of the matter is bound in halos by the end of the EMDE, microhalo formation afterward occurs at most marginally later than $\aeq$.

Our prescription for the impact of gravitational heating will allow explorations of the observational accessibility of EMDE scenarios to be extended to a far wider range of parameters.
For example, a significant portion of the parameter space that Ref.~\cite{hg23} considered in relation to observational probes is impacted by gravitational heating. For scenarios in which the maximum enhancement of density variations would be $\sim 10^{4.4}$, which Ref.~\cite{hg23} took to be the regime where gravitational heating begins to be important, we found that the enhancement is already suppressed by free streaming down to the $\sim 10^2$ level (see Fig.~\ref{fig:peaks}). That outcome makes these scenarios compatible with the Fermi collaboration's measurement of the isotropic gamma-ray background \cite{igrb}, whereas without gravitational heating they would have been ruled out. Gravitational heating is of great relevance to the observational prospects of EMDE scenarios, both via dark matter annihilation and through gravitational detection of microhalos, but we leave a fuller analysis for future work.

A limitation to our results is that we only explored scenarios in which the fraction of matter bound into halos by the end of the EMDE exceeds about 1\%.
Also, for bound fractions around the 10\% level, the form of the cut-off in the power spectrum due to gravitational heating can already begin to differ from the parametrization that is valid for high-bound-fraction cases (see Fig.~\ref{fig:rd4fits}).
Further work is needed to establish the impact of gravitational heating in cases with only marginal structure formation during the EMDE. We remark that this is also the regime in which microhalo formation is expected to begin deep in the radiation-dominated epoch, and further work is separately needed to understand the halos that arise in this way \cite{2023PhRvD.107h3505D}.

\appendix

\section{Calculating the linear velocity dispersion}
\label{linsigv}

This Appendix shows how the velocity dispersion is connected to the power spectrum at linear order in perturbations. Let $\vec x$ be the comoving Lagrangian coordinate. If the linear density contrast field is $\delta(\vec x)$, the comoving displacement field $\vec s(\vec x)$ is related to $\delta(\vec x)$ by $\nabla\cdot\vec s=-\delta$ at linear order. In Fourier space, this implies
\begin{equation}
    \vec s(\vec k)=\frac{\I \vec k}{\vec k^2}\delta(\vec k)
\end{equation}
if $\vec s$ is irrotational. During matter domination, the comoving velocity field is
\begin{equation}
    \dot{\vec s}(\vec k)=H\vec s(\vec k)=H\frac{\I \vec k}{\vec k^2}\delta(\vec k)
\end{equation}
by the Zel'dovich approximation, where $H$ is the Hubble rate. Using the inverse Fourier transform
$\vec s(\vec x)\equiv (2\pi)^{-3}\int\diff^3\vec k\e^{\I\vec k\cdot\vec x}\vec s(\vec k)$,
we write the squared comoving velocity dispersion as
\begin{align}
    \langle [\dot{\vec s}(\vec x)]^2\rangle
    &=
    H^2\int\frac{\diff^3\vec k}{(2\pi)^3}
    \int\frac{\diff^3\vec k^\p}{(2\pi)^3}
    \e^{-\I(\vec k-\vec k^\prime)\cdot\vec x}
    \langle\vec s^*(\vec k)\cdot\vec s(\vec k^\p)\rangle
    \nonumber\\&= H^2
    \int\frac{\diff^3\vec k}{(2\pi)^3}
    \int\frac{\diff^3\vec k^\p}{(2\pi)^3}
    \e^{-\I(\vec k-\vec k^\prime)\cdot\vec x}
    \frac{\vec k\cdot\vec k^\p}{\vec k^2\vec k^{\p 2}}
    \langle\delta^*(\vec k)\delta(\vec k^\p)\rangle.
\end{align}
But $\langle\delta^*(\vec k)\delta(\vec k^\p)\rangle=(2\pi)^3\delta_\mathrm{D}^3(\vec k-\vec k^\p) P(k)$, where $P(k)$ is the power spectrum of $\delta$ and $\delta_\mathrm{D}^3$ is the three-dimensional Dirac delta function. Thus,
\begin{align}
    \langle [\dot{\vec s}(\vec x)]^2\rangle
    &= H^2
    \int\frac{\diff^3\vec k}{(2\pi)^3}
    \frac{P(k)}{k^2}
    = H^2
    \int_0^\infty\frac{\diff k}{k}
    \frac{\mathcal{P}(k)}{k^2},
\end{align}
where $\mathcal{P}(k)\equiv k^3 P(k)/(2\pi^2)$ is the dimensionless power spectrum.
The peculiar velocity dispersion is then
\begin{equation}
    \sigma_{\rm lin}=a\langle [\dot{\vec s}(\vec x)]^2\rangle^{1/2}
    = a H \left(\int_0^\infty\frac{\diff k}{k}
    \frac{\mathcal{P}(k)}{k^2}\right)^{1/2},
\end{equation}
and at reheating it is \begin{equation}
    \sigma_{\rm lin}|_{a=\rh{a}} = 1.04\rh{k} \left( \int_0^\infty \frac{\diff k}{k} \frac{\mathcal{P}(k,\rh{a})}{k^2}\right)^{1/2},
\end{equation} since $(aH)|_{a=\rh{a}} = 1.04\rh{a}\Gamma=1.04\rh{k}$. The numerical factor here follows from the separate definitions of $\rh{T}$ and $\rh{a}$; see Eq.~(\ref{arha0}) and footnote~\ref{foot:H_RH}.

\section{Predicting the Free-Streaming Cut-off Using The Velocity Distribution}
\label{velocity_distribution}

In Section~\ref{cut-off}, we presented a prescription for the free-streaming cut-off that arises from gravitational heating. That prescription, given by Eqs. (\ref{rfseq}) and~(\ref{kfsfit}), was motivated solely by the ratio $\mathcal{P}_\mathrm{sim}(k)/\mathcal{P}_\mathrm{lin}(k)$ of the matter power spectrum in the simulations to that evaluated using linear theory alone. Here we present an alternative prescription, in which we additionally employ information about the velocity distribution in the simulations.

As long as particles cover distances proportional to their streaming velocities, the distribution of particle displacements due to free streaming is just a rescaled version of the velocity distribution.
Consequently, the impact of free streaming is to convolve the density field with this rescaled version of the velocity distribution. This means that -- up to a rescaling in $k$ -- the cut-off function $\rfs$ is the Fourier transform of the velocity distribution.

A more explicit way to see this is to start with the collisionless Boltzmann equation without gravity,
\begin{equation}\label{CBE}
    \frac{\diff f}{\diff t}=\frac{\partial f}{\partial t}+v_i\frac{\partial f}{\partial x_i}=0,
\end{equation}
where $f(\vec x, \vec v, t)$ is the distribution function and the repeated index is summed over. We neglect cosmic expansion for simplicity, but it is easy to check that the conclusion is unchanged if expansion is accounted for. In Fourier space
\begin{equation}\label{CBEk}
    \partial f/\partial t+\I\vec v\cdot\vec k f=0,
\end{equation}
the solution to which is
\begin{equation}\label{fk}
    f(\vec k,\vec v,t) = f(\vec k,\vec v,0)\e^{-\I\vec k\cdot\vec v t}
\end{equation}
for some initial time $t=0$.
Now approximate that the distribution function is separable,
$f(\vec k,\vec v,t) = \rho(\vec k,a) f_v(\vec v)$,
i.e., the velocity distribution $f_v$ is independent of position.
Integrating Eq.~(\ref{fk}) over velocities then yields
\begin{align}\label{fk2}
    \rho(\vec k,t)
    &=
    \rho(\vec k,0)\int\diff^3\vec v f_v(\vec v)\e^{-\I\vec k\cdot\vec v t}
    =
    \rho(\vec k,0)\tilde f_v(\vec k t),
\end{align}
where $\tilde f_v$ is the Fourier transform of $f_v$.
The $k$-space density $\rho(\vec k,t)$ is evidently suppressed, in relation to $\rho(\vec k,0)$, by the Fourier transform of the velocity distribution, scaled in wavenumber by a time-dependent factor. Accounting for cosmic expansion changes the time dependence but otherwise leaves the result unchanged.

\begin{figure}[htb!]
\centering
\includegraphics[width=\textwidth]{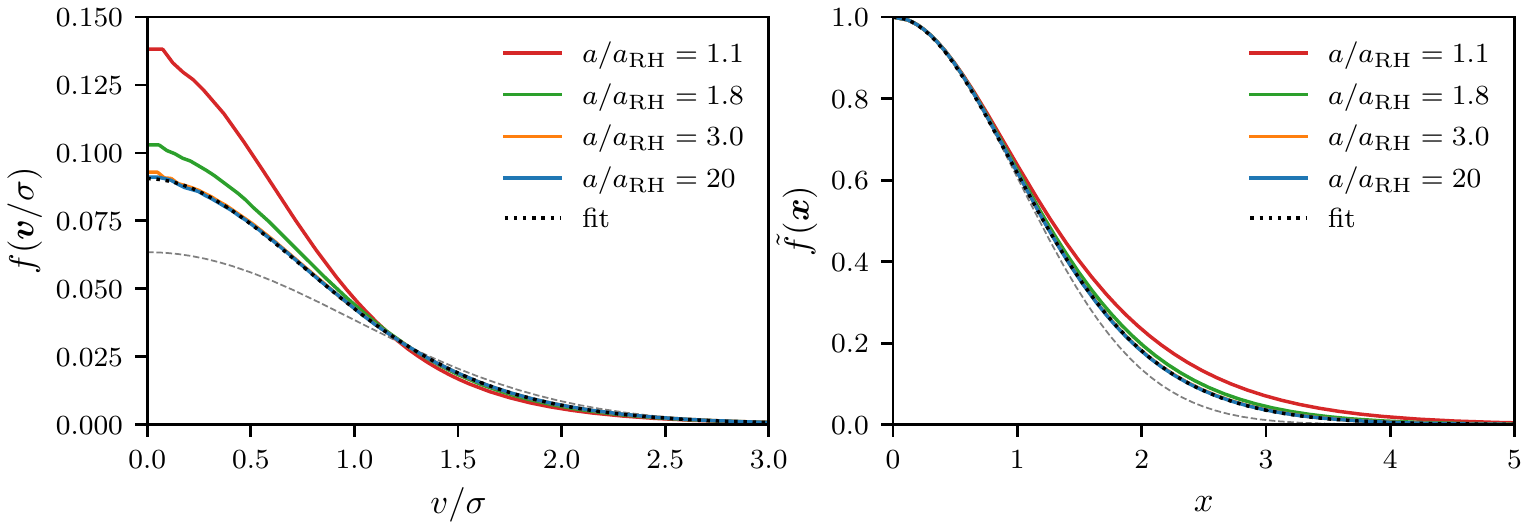}
\caption{The velocity distribution (left) and its Fourier transform (right) in the RD2 simulation at several times (different colors). We show the spherical average and normalize to the velocity dispersion $\sigma=\langle\vec v^2\rangle^{1/2}$. The distribution evidently does not vary after about $3\rh{a}$. We show as black dotted curves the fit to the late-time velocity distribution, given by Eq.~(\ref{vdistfit}), in the left panel, and a fit to its Fourier transform, Eq.~(\ref{vdistFTfit}), in the right panel. For comparison, the gray dashed curves show the Maxwell-Boltzmann distribution.}
\label{fig:velocities}
\end{figure}

The velocity distribution in the RD2 simulation is shown in the left panel of Fig.~\ref{fig:velocities}. The distribution of normalized velocities $\vec v/\sigma$, where $\sigma=\langle\vec v^2\rangle^{1/2}$ is the velocity dispersion, converges quickly after reheating to a form that is well approximated by
\begin{equation}\label{vdistfit}
    f(\vec x)=3.98\times 10^{-11} \left[\e^{(x/14.9)^{1.94}}-0.953\right]^{-7.06}
\end{equation}
(dotted curve), where $\vec x=\vec v/\sigma$. This distribution has a significantly more pronounced low-velocity tail than the Maxwell-Boltzmann distribution (faint dashed curve).

The right panels of Fig.~\ref{fig:velocities} show the three-dimensional Fourier transforms of the velocity distributions in the left panels. At late times, the Fourier transformed distribution is well approximated by
\begin{equation}\label{vdistFTfit}
    \tilde f(\vec x) = \left[2.43\e^{(x/3.69)^{1.99}}-1.43\right]^{-2.81}
\end{equation}
(dotted curve). The strong low-velocity support in Eq.~(\ref{vdistfit}) evidently leads to a long tail in the Fourier transform; $\tilde f$ is significantly larger for $x\gg 1$ than the Fourier transformed Maxwell-Boltzmann distribution (faint dashed curve), which is just a Gaussian function. Per the discussion above, this implies that gravitational heating leads to a longer tail in the cut-off to the power spectrum, in qualitative agreement with the findings in Sec.~\ref{cut-off}.

\begin{figure}[htb!]
\centering
\includegraphics[width=\textwidth]{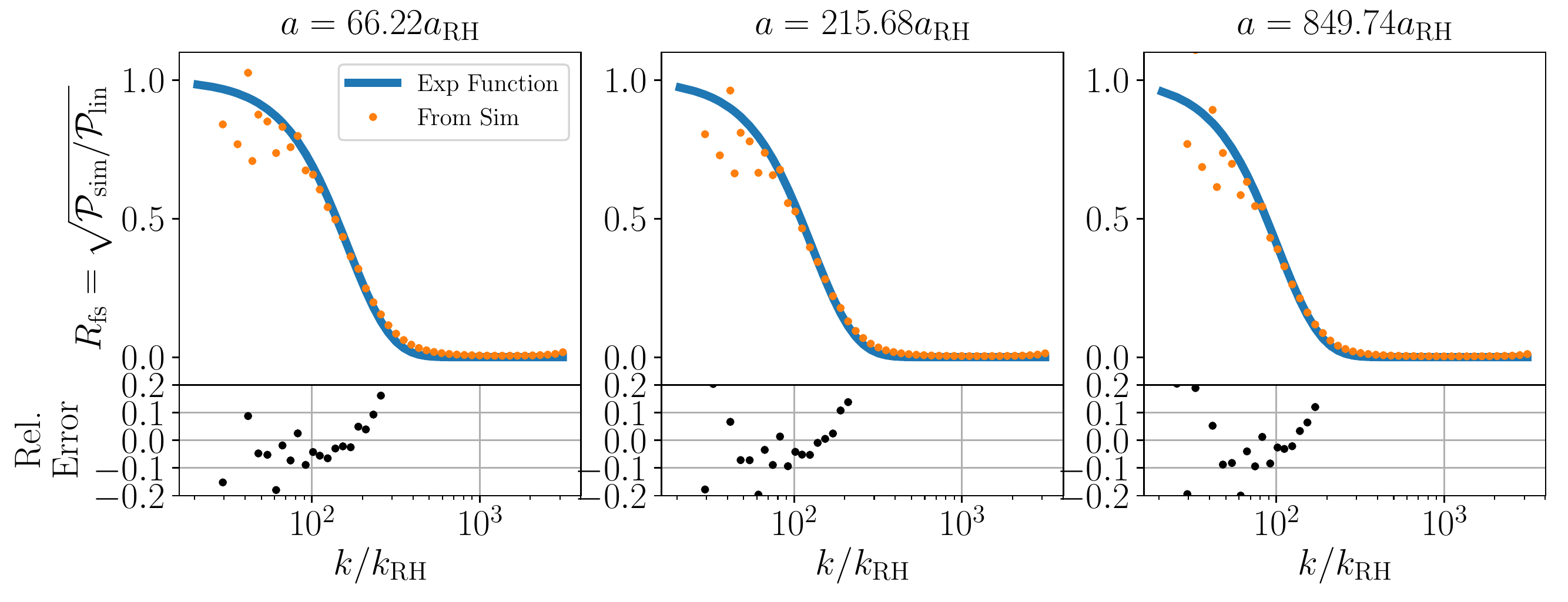}
\caption{$\rfs = \sqrt{\mathcal{P}_{\rm sim}/\mathcal{P}_{\rm lin}}$ from the YD1 simulation box, shown at different times after reheating. The blue curves show the function given by Eq.~(\ref{rfseq2}) with $\kfs$ given by Eq.~(\ref{kfseq2}), while the bottom panels show the relative error between the $\rfs$ from simulations and the blue curves. }
\label{fig:expfit}
\end{figure}

Motivated by Eq.~(\ref{vdistFTfit}), we fit the simulation cut-off $\rfs(k,a)$ (Eq.~\ref{simcut}) to a function of the form \begin{equation} \label{rfseq2}
    \left[2.43\e^{(k/\kfs)^{1.99}}-1.43\right]^{-2.81}.
\end{equation}
We find that the above fitting form works well for the RD1, RD2, RD3, YD1, YD2, YD3, and YD4 simulations. The cut-off scale obtained by this fitting follows the relation \begin{equation} \label{kfseq2}
    \frac{\kfs (a)}{\rh{k}} = \frac{7.41}{\sigma_{\rm t,RH}} \left[ \ln \left( \frac{a}{\rh{a}} \right) \right]^{-1}.
\end{equation}
Figure \ref{fig:expfit} shows the $\rfs$ obtained from the YD1 simulation, with the function given by Eq.~(\ref{rfseq2}) and $\kfs$ given by Eq.~(\ref{kfseq2}). The error between the exponential fitting function and the simulation-based cut-off (lower panel) remains within 10\% for modes for which $\rfs \gtrsim 0.2$. For the RD4 and RD5 cases, the fall-off in $\rfs$ is much shallower than in the other seven cases because of lower structure formation before $\rh{a}$. The prescription in this section does not adequately fit the free-streaming cut-off in these cases. In cases with significant structure formation during the EMDE ($\gtrsim 20\%$), the exponential fit works well.

Compared to the simpler form in Eq.~(\ref{rfseq}), the cut-off function in Eq.~(\ref{rfseq2}) is not an obviously better fit to the simulation results, but it has a stronger theoretical motivation. However, several simplifying assumptions were made in associating the cut-off with the velocity distribution. First, the velocity distribution was assumed to be the same everywhere. In practice, the velocities arise from structure formation and are hence inhomogeneously distributed. Although particles in the high-velocity tail should rapidly homogenize, the low-velocity portion of Fig.~\ref{fig:velocities} (left panel) may include a significant contribution from coherent large-scale motions, which would not suppress power. Additionally, free streaming is assumed to begin instantaneously, whereas in practice the $Y$ particles decay over an extended time interval. These considerations limit the degree to which the velocity distribution can be assumed to predict the correct power spectrum cut-off. For this reason, we continue to adopt in the main text the simpler form obtained in Sec.~\ref{cut-off} from the simulation power spectra alone.

\section{Linear growth function in mixed matter-radiation domination}
\label{growthfac}

Here we derive an expression that is valid long after reheating for the growth of perturbations that entered the horizon before reheating. During the radiation-dominated epoch that follows an EMDE, we can write the general growth function as
\begin{equation}
	D(a)\propto\ln (Ba/\rh{a})
\end{equation}
for subhorizon modes, where $B$ is a constant and $\rh{a}$ is defined as in Eq.~(\ref{arha0}). Note that
\begin{equation}\label{growthcoef}
B=\e^{(\diff \ln D/\diff\ln a)^{-1}}/(a/\rh{a}).
\end{equation}
We numerically evaluate $D(a)$ starting during an arbitrary EMDE at $a=10^{-6}\rh{a}$ and find that the value of $B$ given by Eq.~(\ref{growthcoef}) rapidly settles for $a\gtrsim 4\rh{a}$ to a value of $B\simeq 1.733$. That is, the growth function during the radiation-dominated epoch following an EMDE is
\begin{equation}\label{growth_RD}
	D(a)\propto\ln (1.733 a/\rh{a})
\end{equation}
for modes that are already subhorizon.

At the small scales relevant to EMDE-enhanced structure, baryons are a homogeneous background even during the late matter epoch. Reference~\cite{hu96} derived the linear growth solutions during the transition to late matter domination under this assumption; they are
\begin{equation}\label{growth_solutions}
D_i(a)=(1+a/\aeq)^{\mu_i}
\,_2F_1\!\left(-\mu_i,\frac{1}{2}-\mu_i;\frac{1}{2}-2\mu_i;\frac{1}{1+a/\aeq}\right)
\end{equation}
for $i=1$ and 2, where $_2F_1$ is the hypergeometric function. Here
\begin{equation}
\mu_i=\pm\frac{5}{4}\sqrt{1-\frac{24}{25}\fb}-\frac{1}{4}
\end{equation}
with $+$ and $-$ for $i=1$ and $2$, respectively, where $\fb\simeq 0.157$ is the fraction of the late matter density that is in baryons. Note that when $a\gg\aeq$, the hypergeometric function approaches $1$, so $D_1\propto a^{\mu_1}$ becomes the growing mode and $D_2\propto a^{\mu_2}$ the decaying mode. In the radiation-dominated $a\ll\aeq$ limit, Ref.~\cite{hu96} showed that
\begin{equation}
D_i(a) \to
-\frac{\Gamma(1/2-2\mu_i)}{\Gamma(-\mu_i)\Gamma(1/2-\mu_i)}
\left[
\ln (a/\aeq)+\psi(-\mu_i)+\psi(1/2-\mu_i)-2\psi(1)
\right],
\end{equation}
where $\Gamma(x)$ is the gamma function and $\psi(x)\equiv\diff\ln\Gamma(x)/\diff x$ is the digamma function.

Let us write the growth function as
\begin{equation}\label{growth_combination}
D(a)=A_1 D_1(a)+A_2 D_2(a).
\end{equation}
In the radiation-dominated $a\ll\aeq$ limit, we can match Eqs. (\ref{growth_RD}) and~(\ref{growth_combination}) as well as their first derivatives to obtain
\begin{align}\label{growth_A2}
A_2&=
-A_1
\frac
{\Gamma(-\mu_2)\Gamma(1/2-\mu_2)\Gamma(1/2-2 \mu_1)}
{\Gamma(-\mu_1)\Gamma(1/2-\mu_1)\Gamma (1/2-2\mu_2)}
\nonumber\\&\hphantom{=}\times
\frac
{\ln(1.733 \aeq/\rh{a})-\psi(-\mu_1)-\psi(1/2-\mu_1)+2\psi(1)}
{\ln(1.733 \aeq/\rh{a})-\psi(-\mu_2)-\psi(1/2-\mu_2)+2\psi(1)}.
\end{align}
Meanwhile, if we normalize the growth function so that
\begin{equation}
D(a)\to A_1 (a/\aeq)^{\mu_1}\equiv a^{\mu_1}
\end{equation}
in the matter-dominated $a\gg\aeq$ limit, then
\begin{equation}\label{growth_A1}
A_1=\aeq^{\mu_1}.
\end{equation}
The linear growth function during mixed matter-radiation domination, for modes that entered the horizon prior to the last radiation-dominated epoch, is thus given by the combination Eq.~(\ref{growth_combination}) of the solutions given by Eq.~(\ref{growth_solutions}) with coefficients given by Eqs. (\ref{growth_A1}) and~(\ref{growth_A2}).

\section{Code}

The routines for the time-dependent critical density threshold taken from Ref. \cite{blanco19}, the EMDE transfer functions from Ref. \cite{hg23}, the growth factor used in this paper, and the free-streaming cut-off are available \href{https://github.com/hganjoo/emde_heating}{here}.

\acknowledgments

H.G. is supported by NSF Grant AST-2108931. H.G. thanks Adrienne Erickcek for guiding him through the early stages of this project, and Katherine Mack for her constant support and helpful comments. Simulations for this work were run on the Henry2 / Hazel High Performance Computing Cluster at North Carolina State University. 

\appendix

\bibstyle{unsrt}
\bibliography{refs.bib}

\end{document}